\documentclass[12pt,psf,epsf]{JHEP3}

\usepackage{amssymb}
\usepackage{graphicx}
\usepackage{rotating}

 \newcommand{\be}{\begin{equation}}
\newcommand{\ee}{\end{equation}}
\newcommand{\bea}{\begin{eqnarray}}
\newcommand{\eea}{\end{eqnarray}}
\def\NP{ Nucl. Phys. }
\def\PL{ Phys. Lett. }
\def\PR{ Phys. Rev. }

\def\CQG {Class. Quant. Grav.}

\title{Critical Trapped Surfaces Formation in the Collision of
Ultrarelativistic Charges
 in  $(A)dS$}
\author{I.Ya. Aref'eva$^1$,   A.A.Bagrov$^1$ and L.V.  Joukovskaya$^2$ \\
 $^1$\small{Steklov Mathematical Institute, Russian Academy of Sciences,}\\
\small{Gubkina st. 8, 119991, Moscow, Russia}
 \\
 $^2$\small{Centre for Theoretical Cosmology, DAMTP, CMS, University of
Cambridge},\\ \small{ Wilberforce Road, CB3 0WA, Cambridge, United Kingdom} }

\abstract{ We study the formation of marginally trapped surfaces
in the head-on collision of two ultrarelativistic charges in
$(A)dS$ space-time. The metric  of ultrarelativistic charged
particles in $(A)dS$ is  obtained by boosting
Reissner-Nordstr\"om  $(A)dS$  space-time to the speed of light. We
show that formation of trapped surfaces on the past light cone is only possible when
charge is below certain critical - situation similar to the
collision of two ultrarelativistic charges in Minkowski
space-time. This critical value  depends on the energy of
colliding particles  and the value of a cosmological constant.
There is richer structure of critical domains in $dS$ case. In
this case already for chargeless particles there is a critical
value of the cosmological constant only below which trapped
surfaces formation is possible. Appearance of arbitrary
small nonzero charge significantly changes the physical picture.
Critical effect which has been observed in the neutral case does
not take place more. If the value of the charge is not very large
solution to the equation on trapped surface exists for any values
of cosmological radius and energy density of shock waves.
Increasing of the charge leads to decrease of the trapped surface
area, and at some critical point the formation of trapped surfaces of the type mentioned above
becomes impossible.

}


\begin{document}

\newpage

\section{Introduction}

Black holes are expected to form in collisions of
ultrarelativistic particles with energies above the
Planck scale \cite{thooft_s,Dray,ACV,Peter,AVV}. The Planck energy could be
few TeV in the framework of TeV-gravity, where   our space is
a 3-brane situated in a large extra dimensional space  and
elementary particles  are confined on the brane~\cite{TeVgravity}.
In this scenario the black hole production in collisions of particles with the center-mass energy of a few TeV
and their experimental signatures \cite{TeVgravity-HEC}  at the LHC
became the subject of intensive
 analytical and numerical investigations
\cite{BF,IA,DL,GT,EG02,BH-list1,YN03,YR05,num}. We also note a discussion of
the possible production of wormholes and others more exotic objects at
the LHC \cite{TM,MMT,NS}. See \cite{INov,NovNov} for a consideration
of wormholes in astrophysics.

There are  motivations to study  similar processes in $AdS$
background. In the framework of $AdS/CFT$ correspondence
\cite{adscft} the
 formation of  trapped surfaces in $AdS_5 $ is interesting because it is supposed
 to be dual to real four-dimensional  formation of  the quark-gluon plasma and
 its  thermal equilibration \cite{Gubser08,Alvarez08,Gubser09,
 0803.1467,Shuryak08,0803.3226,
Shuryak09,0902.3046,0904.2536,0906.4426,0907.4604,0908.3677}
Existence of a critical impact parameter, beyond which the
trapped surface does not exist has been considered as an  indication for
similar critical
impact parameter in real collisions of heavy ions at the Relativistic Heavy Ion
Collider (RHIC)
at Brookhaven National Lab.
 We would like to recall that applications of the
$AdS/CFT$ correspondence  to strongly coupled quark-gluon plasma
 has lead to many interesting  results
\cite{Shuryak08,Tseytlin,Starinets,Solana,Zahed}, but these
results are related to description equilibrium   or
 close to equilibrium quark-gluon plasma.

There are   observational as well as theoretical reasons to
consider collisions of ultrarelativistic particles  in
asymptotically $dS_4$ space-times. Well-known experimental fact is that
our universe is expanding with constant acceleration which is well described
 by a small positive
cosmological constant and this  implies that
the space-time is asymptotically $dS$. Furthermore  processes of collisions
of a ultrarelativistic matter could be
 also interesting in the context of the early Universe \cite{Brill:1993tw,Romans:1991nq,ross,bousso,Jolien,Kastor}.
 For  the current status of black hole astrophysics see \cite{Narayan:2005ie}.

Let us note that trapped surface formation  in the collision of
ultra-relativistic particles
in dS space-time could be interesting also
 in the context of
dS/CFT correspondence  \cite{Hull98,Hull98signature,Minic01,Witten01ds-dul,Strominger01ds-dul}.
\par

It is known that  in the   flat background \cite{YosMann,Hatta-charge}
charges of ultrarelativistic
colliding
particles (for  a fixed impact parameter) decrease the trapped surface area and
by increasing the charges
one  may reach  a critical point at which no trapped surfaces can be formed.
It is natural to ask a  similar question
 about  critical effects related to charges of ultrarelativistic
 particles propagating   in $(A)dS$ background.

In this paper the formation of trapped surfaces in collisions
of two charged shock waves in $(A)dS$ background is being considered.
The  first work in this direction in the $AdS/CFT$ context  was the paper by
 Gubser, Pufu and Yarom \cite{Gubser08},  who considered central collisions of
neutral  ultrarelativistic particles and found the
(marginally) trapped surface forming at the collision moment.
The  trapped surface area  was then used as  a lower
bound estimation of the entropy produced in a heavy ion collision.
In the limit of very large collision energy $E$ they found that the
entropy grows as $E^{2/3}$.

Lin and Shuryak \cite{Shuryak09} have discussed matching heavy ion collisions to those of gravitational
shock waves
and
 have noted that  the effective size of objects in gravitational
 collision grows with energy with an exponent ten times that in the real QCD.
 They have stressed  that
one  cannot  tune  the scale of the cosmological constant $\Lambda$ or the impact parameter
 of a bulk colliding objects to the size of nucleus and tune it perhaps to the parton density.
Non-zero  impact parameter case is considered in \cite{Gubser09,Shuryak09}
and a natural critical phenomenon
 analogous to neutral shock wave collision in Minkowski background
\cite{giddings,nambu,veneziano}  has been found. Beyond certain impact parameter,
the trapped surface disappears and black hole formation
does not happen.

Other type of criticality has been found by  Alvarez-Gaume,
Gomez, Vera, Tavanfar and Vazquez-Mozo
\cite{Alvarez08}.
They have considered
 central collision of shock waves sourced by a matter
distribution in the transverse space and have found  critical
phenomenon occurring where  the shock
wave reaches some diluteness limit and the formation of the marginally
trapped surface is no longer possible.

In this paper we  find a new  critical phenomenon in $AdS$ background.
Namely, in the $AdS$ case   beyond certain charge,
the trapped surface disappears and black hole formation
does not happen.  The value of the critical charge depends on the energy of colliding particles
as well as the value of the cosmological constant.
This phenomenon  is analogous to the  critical
phenomenon found by Mann and Yoshino in
the charged shock wave collision in Minkowski background
\cite{YosMann}.
 We consider here head-on collisions only and no-head ones can be considered by
straightforward  generalization of the technics developed in
\cite{Gubser09} and one can expect that there exists the domain in
$(Q,b)$-plane beyond which the trapped surface disappears
(here $Q$ is charge and $b$ is impact parameter).

There is a richer structure of critical domains in the $dS$ case.
 In this case already in absence of a charge there is a critical
  value for the ratio of the shock wave energy density
  and the cosmological radius only below which trapped surfaces formation
  is possible \cite{ABG}. We observe that
   small charge substantially violates this critical effect.
   If the charge is nonzero but enough small the trapped surface on the past light cone can be formed
   for given values of cosmological constant and energy density of shock wave
   even if in the neutral case for the same parameters the trapped surface can not be formed.
   Increasing of the charge decrease the area of formed trapped
   surface and it can be produced until
 the charge does not overcome  critical value which  depends on the energy
 of colliding particles  and the value of the cosmological constant.

The paper is organized as follows. In Sect.2.1. we present the $(A)dS$ analog of
the charged  Aichelburg-Sexl shock wave   which describes charged ultrarelativistic
particles in the (A)dS space time. This is a generalization of results of
Hotta and Tanaka \cite{HT} to the case of charged particles. We note a problem that appears in the $dS$ case
and that is associated  with  the cosmological horizon of $dS$ metric and can be solved by regularization procedure.
Physical meaning of this regularization requires more detailed  investigations.
In Sect.2.2 we present the standard picture of two ultrarelativistic
colliding particles and the trapped surface equation for the central
collision in the the $(A)dS$ case. Sect. 3 is devoted to description of the solutions
to equation which defines a radius of trapped surface for the central collision in the
$AdS_5$ case. The existence of the critical charge is being demonstrated.
In Sect. 4 we discuss solutions to the trapped surface equation in  the $dS_4$
case.

\section{Set up}
\subsection{Metric of an ultrarelativistic charge in  the (A)dS background}
The metric of  an ultrarelativistic charge in the flat background
has been obtained \cite{LS90,Ortaggio06} by applying the
Aichelburg-Sexl boost \cite{AS71}
 to $D$-dimensional Reissner-Nortstr\"om space-time \cite{MP86}, see also
 \cite{Gibbons-Kerr}.

 We apply the same procedure to Reissner-Nordstr\"om-(anti)de Sitter black
 hole metric. Our calculations generalize results of Hotta and Tanaka \cite{HT}, see also
 \cite{LS90,Sfetsos,GrifPod,Emparan,Ortaggio06,Eso,Nastase1,Nastase2,Nastase3}.
In a spherical static (Schwarzschild) coordinates the RN-$(A)dS$ metric
 has the form
(see for example, \cite{Radu})
\begin{equation}
\label{RN-lambda}
ds^2=-g(R)dT^2+g(R)^{-1}dR^2+R^2d\Omega_{D-2}^2,
\end{equation}
where
\bea
g(R)=1-\frac{2M}{R^{D-3}}+\frac{Q^2}{R^{2(D-3)}}\pm\frac{\Lambda}{3}R^2,\eea
and the electromagnetic field
\bea
A=A_TdT=\Big(\sqrt{\frac{D-2}{2(D-3)}}\frac{ Q}{R^{D-3}}+\Phi \Big) dT,
\eea
gives a solution to the E.O.M.  for a (pure electric) gauge
potential, here $\Phi$ is a constant and $\Lambda$ is a cosmological constant,
 $\Lambda/3\equiv 1/a^2$.

Here and below $(-)$ corresponds to  $dS$  and  $(+)$ to  $AdS$ .
$Q$ and $M$ are related to charge $q$ and mass $m$
\begin{equation}
Q^2=\frac{8\pi G_Dq^2}{(D-2)(D-3)},
\end{equation}
\begin{equation}
M=\frac{8\pi G_Dm}{(D-2)\Omega_{D-2}}.
\end{equation}

To get the metric of an ultrarelativistic charge in $(A)dS$
we
take metric (\ref{RN-lambda})   and expand  it on $M$ and $Q^2$. Thereafter
we take the first order terms on these parameters
\be
\label{NN-lambda}
ds^2=ds^2_{dS,AdS}+(\frac{2M}{R^{D-3}}-\frac{Q^2}{R^{2(D-3)}})dT^2+
\frac{1}{(1\mp\frac{R^2}{a^2})^2}(\frac{2M}{R^{D-3}}-\frac{Q^2}{R^{2(D-3)}})dR^2
\ee
We rewrite (\ref{NN-lambda}) in the plane coordinates which are satisfying relations
\bea
   \label{ds}            -Z_{0}^{2}+Z_{1}^{2}+Z_{2}^{2}+...+Z_{D-1}^{2}+Z_{D}^{2}=a^{2},\\
                \label{ads} -Z_{0}^{2}+Z_{1}^{2}+Z_{2}^{2}+...+Z_{D-1}^{2}-Z_{D}^{2}=-a^{2},
  \eea
  Plane coordinates are related to $T$ and $R$ via formula
\bea
Z_0&=&\sqrt{a^2-R^2}\sinh T/a\\
Z_{D}&=&\pm\sqrt{a^2-R^2}\cosh T/a
\eea
    for $dS$ and    \bea
            Z_{0} &\equiv& \sqrt{a^{2}+R^{2}} \; \sin(T/a),
            \label{ad-Z0}
            \\
            Z_{D} &\equiv& \sqrt{a^{2}+R^{2}} \; \cos(T/a),
            \label{ad-ZD}
            \eea
 for  $AdS$ case and
\bea
Z_1&=&R\cos \theta_1,\\
Z_2&=&R\sin \theta_1\cos \theta_2,\,...,\,\\
Z_{D-1}&=&R\sin \theta_1\sin\theta_2...\sin\theta_{D-2},
\eea
for both cases.  As a result we have
\be
    ds^{2}= ds_{0}^{2}+ds_{p}^{2},
    \ee
where $ds_{0}^{2}$ is $(A)dS$ metric and the perturbation $ds_{p}^{2}$ has the form
\bea
    ds_{p}^{2}=G_{00}dZ^2_0
    +G_{DD}dZ_D^{2}+
    G_{0D}dZ_0dZ_D,
\end{eqnarray}
where
\be
G_{MN}=\chi(Z_0^2,Z^2_D,M,Q)\cdot g_{MN}(Z_0^2,Z^2_D)
\ee
 and  nonzero components of $g_{MN}$ and the overall factor
$\chi(Z_0^2,Z^2_D,M,Q)$ are given by
\bea
g_{00}&=&\pm a^2Z^2_D+Z_0^2Z^2_D\mp Z_D^4+ a^2Z^2_0,\\
g_{DD}&=&\pm a^2Z^2_0+Z_0^4\mp Z_D^2Z^2_0+ a^2Z^2_D.\\
g_{0D} &=&- 2(\pm 2a^2+Z_0^2\mp Z_D^2)Z_0Z_D,\\
\chi &=& \frac{a^2 }{(Z_D^2\mp Z^2_0)^2}\left(\frac{2M}{(\pm a^2+Z_0^2\mp Z^2_D)^{\frac{D-1}{2}}}-
\frac{Q^2}{( \pm a^2+Z_0^2\mp Z^2_D)^{D-2}}\right).
   \eea

    Performing  a boost in the $Z_{1}$-direction
\bea
    Z_{0}&=&\gamma (Y_{0}+v Y_{1}),\,\,\,\,\,\gamma \equiv (1-v^{2})^{-1/2},
    \label{(19)}
\\
    Z_{1}&=&\gamma (vY_{0}+Y_{1}),
    \label{(20)}
\\
    Z_{2}&=&Y_{2}, \,... \; Z_{D}=Y_{D}.
    \label{(21)}
\eea
and rescaling
\bea
    M&=&\bar{M}\sqrt{1-v^2}\equiv \bar{M}/\gamma,\\
    Q^2&=&\bar{Q}^2\sqrt{1-v^2}\equiv \bar{Q}^2/\gamma,
\eea
  we put the first order deformation of  the metric into the form
\bea
 ds_{p}^{2}
  &=&\gamma G_{00}(\gamma^2 (Y_{0}+v Y_{1})^2,Y_D^2)\, d (Y_{0}+v Y_{1})^2\nonumber\\
  &+&
    G_{DD}(\gamma^2 (Y_{0}+v Y_{1})^2,Y_D^2)\,d (Y_{0}+v Y_{1})\, dY_D\nonumber\\
    &+&
   \frac1\gamma G_{DD}(\gamma^2 (Y_{0}+v Y_{1})^2,Y_D^2)\,dY_D^{2}.
   \label{dsp}
      \eea

To get the limit when $\gamma \to \infty$ in the case of $AdS$ like in
absence of a charge we can  apply  the lemma  of \cite{AS71,HT}
that is dealing  with distributions:

\be
    \label{lim}
    \lim _{v\to 1}\gamma f\left(\gamma^2 (Y_{0}+v Y_{1})^2\right)=\delta(Y_{0}+ Y_{1})
    \int f(x^2)dx
\ee

The case of $dS$ is more subtle. The problem is that
 in the $dS$ case components of  $ds_{p}^{2}$
  are not distributions for any value of $D$.
 Indeed, we need to consider the following expressions
\be
\label{lsin}
 \frac{\gamma }{(Z^2-\gamma^2Y^2)^2}f(\gamma^2 Y^2),
\ee
where $f$ is a smooth function.

It is well known \cite{Gel,VSV} that  $\frac{1 }{(Z^2-\gamma^2Y^2)^2}$
(as function of $Y$)
is not  a distribution for any $\gamma\neq 0$.  To treat the expression (\ref{lsin})
as a distribution one has to use a regularization.
  All natural regularizations have been
studied \cite{Gel,VSV} and we would    use one of them.
We can construct modification of the formula (\ref{lim}) which
operates with
\be
\label{lim-sin}
 \lim _{\gamma\to \infty}\left(\gamma \left(\frac{1}{(Z^2-\gamma^2Y^2)^2}\right)_{reg}
 f(\gamma^2 Y^2)\right)
\ee
and is suitable for the $dS$ case (see Appendix A.2):
\bea
\label{lim-sin-l}
\lim _{\gamma\to \infty}\left(\frac{\gamma }{(Z^2-\gamma^2Y^2)^2}\right)_{reg}f(\gamma^2 Y^2)
 =\delta(Y)\int \left(\frac{1}{(Z^2-x^2)^2}\right)_{reg}f( x^2)dx
\eea

According to mentioned formulas
we have  to select only terms proportional
to $\gamma$ ($\gamma\to \infty$ when $v \to 1$).
This prescription
gives the following answer
\be
\label{ans}
ds^2=ds^2_{0}+F_{D,dS/AdS}(\bar{M},\bar{Q}^2,Y_D)\delta(u)du^2,
\ee
where $u=Y_{0}+ Y_{1}$ , and the shape function
$F_{D,dS/AdS}(\bar{M},\bar{Q}^2,Z)$ is given by the following formula
\be
     \label{final-universal}
 F_{D,dS/AdS}(\bar{M},\bar{Q}^2,Z)=F_{D,dS/AdS}(\bar{M},Z)-
 \frac{\bar{Q}^2}{2\bar{M}}F_{2D-3,dS/AdS}(\bar{M},Z),
 \ee
 where
 \bea F_{D,dS/AdS}(\bar{M},Z)  &=&2\bar{M}a^2\int_{-\infty}^{\infty}\left[
    \frac{\left( a^2 (\pm Z^2+ x^2)
    +Z^2(x^2\mp Z^2)\right) }{( Z^2\mp x^2)^2\cdot
    ( \pm a^2+x^2\mp  Z^2)^{\frac{D-1}{2}}}\right]dx.
    \label{ID-dS}
   \eea

Therefore,
$F_{D,dS/AdS}(\bar{M},\bar{Q}^2,Z_D)$ is a sum of the profile function for
     the chargeless shock wave  in the same space-time dimension plus the profile function of
     $2D-3$  dimensional  chargeless shock wave
     multiplied  by a ratio of the charge  square to  mass $\bar{M}$.

\subsection{Two waves picture and  trapped surface equations}
\begin{figure}[h]
    \begin{center}
\includegraphics[height=5cm]{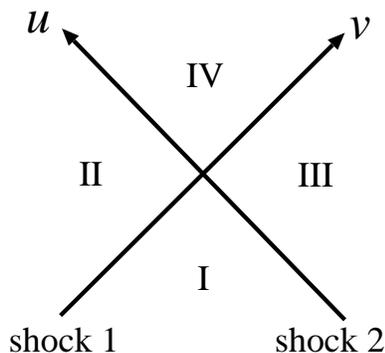}
\end{center}
\caption{Schematics  picture of two colliding ultrarelativistic particles.
$u$ and $v$ is a set of the light-cone coordinates
}\label{fig:2shock}
\end{figure}

In the previous subsection we presented gravitational properties of one
ultrarelativistic charged particle
 traveling in $(A)dS$  space-time.
The gravitational field of the
 particle is infinitely Lorentz-contracted and forms a
shock wave.  Except at the shock wave, the space-time is $(A)dS$. To
deal with two colliding ultrarelativistic particles one deals with
picture schematically shown in Fig.\ref{fig:2shock}. There are two
shock waves  and except at the shock waves the space-time is $(A)dS$
before the collision (i.e., regions I, II, and III). After the
collision these two shocks nonlinearly interact with each other and
the space-time within the future lightcone of the collision (i.e.,
region IV) becomes highly curved. It is unknown how to derive  the
metric in region IV even numerically. But it is
  possible to investigate the apparent horizon  on the slice $u\le 0=v$
and $v\le 0=u$ and calculate the cross section for the apparent
horizon formation $\sigma_{\rm AH}$ in $AdS$ and $dS$ cases, see
\cite{Gubser08} and \cite{ABG}, respectively, similarly to the
flat case \cite{EG02,BH-list1,YN03}.

  In  non asymptotically flat cases one has no  general theorems that
  guarantee formation of
  a black hole if the   marginally trapped surface is formed
  \cite{HawkingPenrose,Penrose}. However
   there is a common opinion
that the existence of the marginally trapped surface can be used as an
indication of
a black hole formation and the area of the trapped surface can be used as a low
bound to estimate
the cross-section of the black hole formation \cite{Gibbons,0001003,0004032,0803.2526}.
The problem of horizon formation in asymptotically non-flat spacetimes has been considered in \cite{hoop}.
\par
This picture  generalizes  the picture for ultrarelativistic particles
without  charge and there is nothing special in the
charged case except the different form of shape functions for the shock waves.

According to this picture the trapped surface is made up of two pieces,
each of them is associated with one of two
 shock waves and  one has to solve the boundary problem for two functions
 that define shapes of these two pieces of the   trapped surface.
 For a central collision these two  functions  are reduced to one
 trapped surface function $\Psi$
  and one
 deals with one equation.
 The form of this equation in the flat, AdS and dS spaces
 has been obtained
  by Eardley and Giddings \cite{giddings}, Gubser, Pufu and
 Yarom \cite{Gubser08} and
in \cite{ABG}, respectively. Corresponding trapped surface
functions  are related to solutions to the second order
differential equation defining eigenfunction problem for the
Beltrami-Laplace operators on
 the plane, Lobachevski space $\mathbb{H}^{D-2}$, and  the sphere
 $\mathbb{S}^{D-2}$ for the flat, AdS, and dS cases
 respectively. The corresponding equations in last two cases have the form

\bea
\left(
\Delta_{\mathbb{S}^{D-2}/\mathbb{H}^{D-2}}\pm\frac{D-2}{a^2}
\right)
\left( \frac{\Psi-\kappa H}{1\pm \frac{\rho^2}{2a^2}} \right)=0,\\
H =\frac12 \left( 1\pm \frac{\rho^2}{2a^2}\right) F \eea where the
upper sign corresponds to the $dS$ case and the lower one to the
$AdS$ case; $\Psi$ is the trapped surface function and $H$ is
related to the shape of shock wave. Note that this
form of the trapped surface equation does not depend on a
particular choice of the shape $F$ of shock wave. Here
$\kappa=\theta(0)$ and we suppose that Heaviside function
$\theta(0)=\frac12$.\par

In the $(A)dS$ case
  it is convenient to work with a chordal distance $q$, that is related to the
  plane coordinates $Z_D$ via
  \be
\label{cordal-AdS} q_{AdS}=\frac{Z_D}{2a}-\frac12
\ee
in the $AdS$ case
and
\be
\label{cordal-dS}
q_{dS}=\frac{1}{2}-\frac{Z_D}{2a}
\ee
in the $dS$ case.

Also we use here radial coordinate $\rho$ related to the chordal coordinates via
\be
    \rho_{AdS}=a\sqrt{\frac{2q_{AdS}}{1+q_{AdS}} },\label{q-rho-AdS}
    \ee
    and
    \be
   \label{q-rho-dS}
    \rho_{dS}=a\sqrt{\frac{2q_{dS}}{1-q_{dS}} }.
    \ee

    The trapped surface function  $\Psi$ for head-on collisions
    has to satisfy
the following boundary conditions on a submanifold $\rho=\rho_0$:
\bea
\Psi|_{\rho=\rho_0}&=&0,\\
\partial_{\rho} \Psi|_{\rho=\rho_0} &=&- 2.
\eea

 Existence of the trapped surface means the existence of
a real solution to   the following equation
  \be
    \label{rts-ads}
    \frac14
(1-\frac{\rho_{0}^2}{2a^2})F'(\rho_{0})-\frac{\rho_{0}}{2a^2+\rho_{0}^2}F(\rho_{0})+\frac{\sqrt{2}}{2\kappa}=0,
\ee
for the $AdS$ case \cite{Gubser08} and

\be
    \label{rts-ds}
    \frac14
(1+\frac{\rho_{0}^2}{2a^2})F'(\rho_{0})+\frac{\rho_{0}}{2a^2-\rho_0^2}F(\rho_{0})+\frac{\sqrt{2}}{2\kappa}=0,
\ee
for the $dS$ case \cite{ABG}. $\rho_0$ defines the radius of the corresponding  trapped surface.
\par One can rewrite equations (\ref{rts-ads}) and (\ref{rts-ds}) in terms of $q$

\be
\label{qc-eq-m}
F^\prime (q_0)-\frac{2}{1+2q_0}F(q_0)+\frac{\frac{2a}{\kappa}}{\sqrt{q_0(1+q_0)}}=0
\ee
for the AdS case and

  \be
\label{eq-q-dS}F^\prime(q_0)+ \frac{2}{1-2q_0}F(q_0)+ \frac{\frac{2a}{\kappa}}{\sqrt{q_0(1-q_0)}}=0.
\ee
for dS case.

 \par  In \cite{Gubser08} $\kappa=1$ and in \cite{ABG} $\kappa=1/2$.
Note that in \cite{ABG} we used different normalization
$F_{[56]}=\sqrt{2}F$.

These equations  determine
values of critical charges for the
trapped surfaces formation. In the next two sections we are going to analyze particular
cases of these equations.

\section{Solution to Trapped Surface Equation in $AdS_5$}

\subsection{Critical charge}

  According to (\ref{final-universal}) the profile of the charged
shock wave in  the $AdS_5$ background has the form
\bea
&\,&F_{5,AdS}(\bar{M},\bar{Q}^2,Z_5) = \\
&=& \nonumber -\frac{3\pi \bar{M}}{a}\,\left(\frac{1
-\frac{2Z_5^2}{a^2}}{ \sqrt{\frac{Z_5^2}{a^2}-1}} +
\frac{2Z_5}{a}\right) -\frac{5\pi
\bar{Q}^2}{8a^3}\frac{12\frac{Z_5^2}{a^2}-3-8\frac{Z_5^4}{a^4}+
\sqrt{\frac{Z_5^2}{a^2}-1}+
8\frac{Z_5}{a}(\frac{Z_5^2}{a^2}-1)^{\frac32}}{(\frac{Z_5^2}{a^2}-1)^{\frac32}},\eea
or in terms of the chordal coordinate $q$

\bea
\label{AdS-prof-mQ-r}
&\,&F_{5,AdS}(\bar{M},\bar{Q}^2,Z_5(q))=\frac{3\pi \bar{M}}{a}\,
\left(\frac{(8q^2+8q+1)-4(2q+1)\sqrt{q(1+q)}}{ 2\sqrt{q(1+q)}}\right)\\
&+&\frac{5\pi \bar{Q}^2}{64a^3}\,\left(\frac{144q^2+16q-1+128q^4+256q^3
-64(2q+1)q(q+1) \sqrt{q(1+q)}}{q(q+1)\sqrt{q(1+q)}}\right).\nonumber
\eea

\begin{figure}[h]
    \begin{center}
\includegraphics[width=5.5cm]{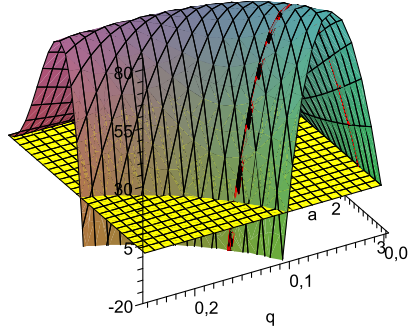}
\end{center}
   \caption{
   $-{\cal N}_{5,AdS}(a,\bar{M},q)$ as function of two variables $q$
   and $a$, $\bar{M}=1$. The yellow plane corresponds to a given value of $15 \frac{ \pi}{a}\,\bar{Q}^2$ } \label{picture-crit-AdS}\end{figure}

\begin{figure}[h]
    \begin{center}
    A.\includegraphics[width=6cm]{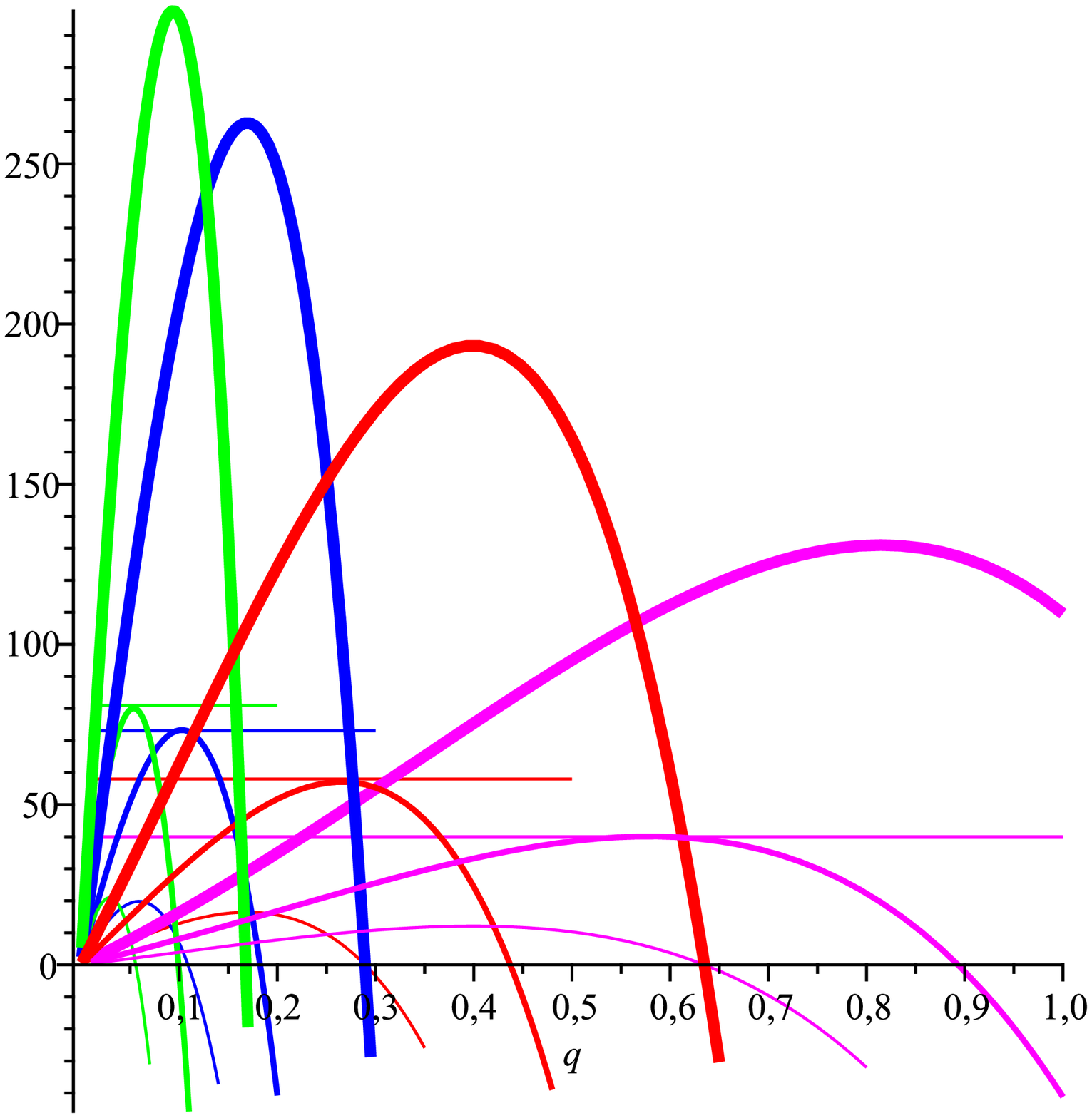} $\,\,\,\,\,\,\,$
B.
\includegraphics[height=6cm]{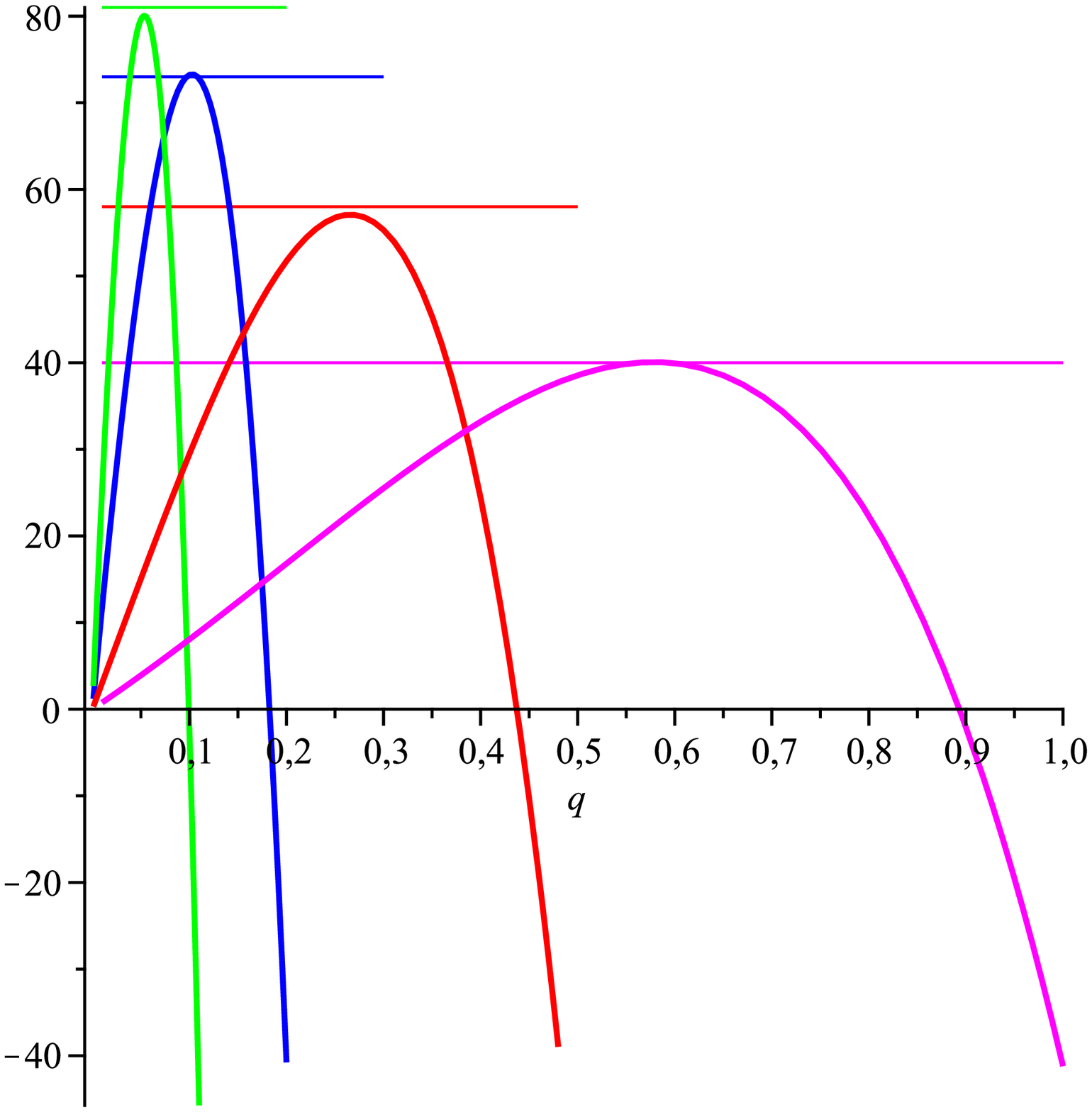}
\end{center}
\caption{
A. $-{\cal N}_{5,AdS}(a,\bar{M},q)$ as function of $q$
   for different values of $a$ and different values of $\bar{M}$.  Thick lines correspond to shock waves with higher
    energy. Magenta lines correspond to $a=0.5$, red lines to $a=1$,
   blue lines to $a=2$
   and green lines to $a=3$. Critical values of the charge increase with increasing of the cosmological radius
   as well as the energy of shock waves. B. $-{\cal N}_{5,AdS}(a,\bar{M},q)$ as function of $q$
   for different values of $a$ and $\bar{M}=1$. The horizontal lines show the critical
   values of $\bar{Q}$ for the correspondent value of $a$. Magenta lines correspond to $a=0.5$, red lines to $a=1$,
   blue lines to $a=2$
   and green lines to $a=3$  and critical values of the charge increase with increasing of the cosmological radius.
 }
\label{picture-crit-AdS-0}
\end{figure}

To find solutions to (\ref{qc-eq-m}) for $F$ given by
(\ref{AdS-prof-mQ-r}) we present the LHS of (\ref{qc-eq-m}) for $F$ given by
(\ref{AdS-prof-mQ-r})

\be
{\cal F}_{5,AdS}(a,\bar{M},\bar{Q}^2,q))\equiv
F^\prime _{5,AdS}(\bar{M},\bar{Q}^2,q)-\frac{2F_{5,AdS}(\bar{M},\bar{Q}^2,q)}{1+2q}+
\frac{2a}{\sqrt{q(1+q)}},
\ee
as
\be
\label{d-na}
{\cal F}_{5,AdS}(a,\bar{M},\bar{Q}^2,q)=
\frac{{\cal N}_{5,AdS}(a,\bar{M},\bar{Q}^2,q)}{{\cal D}_{5,AdS}(a,q)}.
\ee
 The  numerator ${\cal N}_{5,AdS}(a,\bar{M},\bar{Q}^2,q)$
contains just one  term
with dependence on $\bar{Q}^2$. This dependence  is linear
with a positive coefficient
\be
{\cal N}_{5,AdS}(a,\bar{M},\bar{Q}^2,q)={\cal N}_{5,AdS}(a,\bar{M},q)+
15 \frac{ \pi}{a}\,\bar{Q}^2.
\ee
 The denominator in (\ref{d-na})  does not take infinite values. To find solutions to  (\ref{qc-eq-m})
for the shape function given by (\ref{AdS-prof-mQ-r}) we can draw the function
\bea
\nonumber
-{\cal N}_{5,AdS}(a,\bar{M},q)&\equiv& -(512a^3 q^5+1280 a^3 q^4-
96 \bar{M}\pi a q^2+1024 a^3 q^3-96\bar{M} \pi a q\\&+&256 a^3 q^2),
\eea
and  see where this function  is positive and where it can be equal
 to a given value
$
15 \bar{Q}^2\frac{ \pi}{a}.
$
The maximum values of the function $-{\cal N}_{5,AdS}(a,\bar{M},q))$
define the critical values of $\bar{Q}^2$ for given $a$ and $\bar{M}$, see Fig.\ref{picture-crit-AdS}.
In \ref{picture-crit-AdS-0} we present the same pictures for different fixed values of $a$ and $\bar{M}$.

$$\,$$
$$\,$$
$$\,$$
$$\,$$

\subsection{Area of the trapped surface in  $AdS_5$}

\begin{figure}[h]
    \begin{center}
    A.\includegraphics[width=6cm]{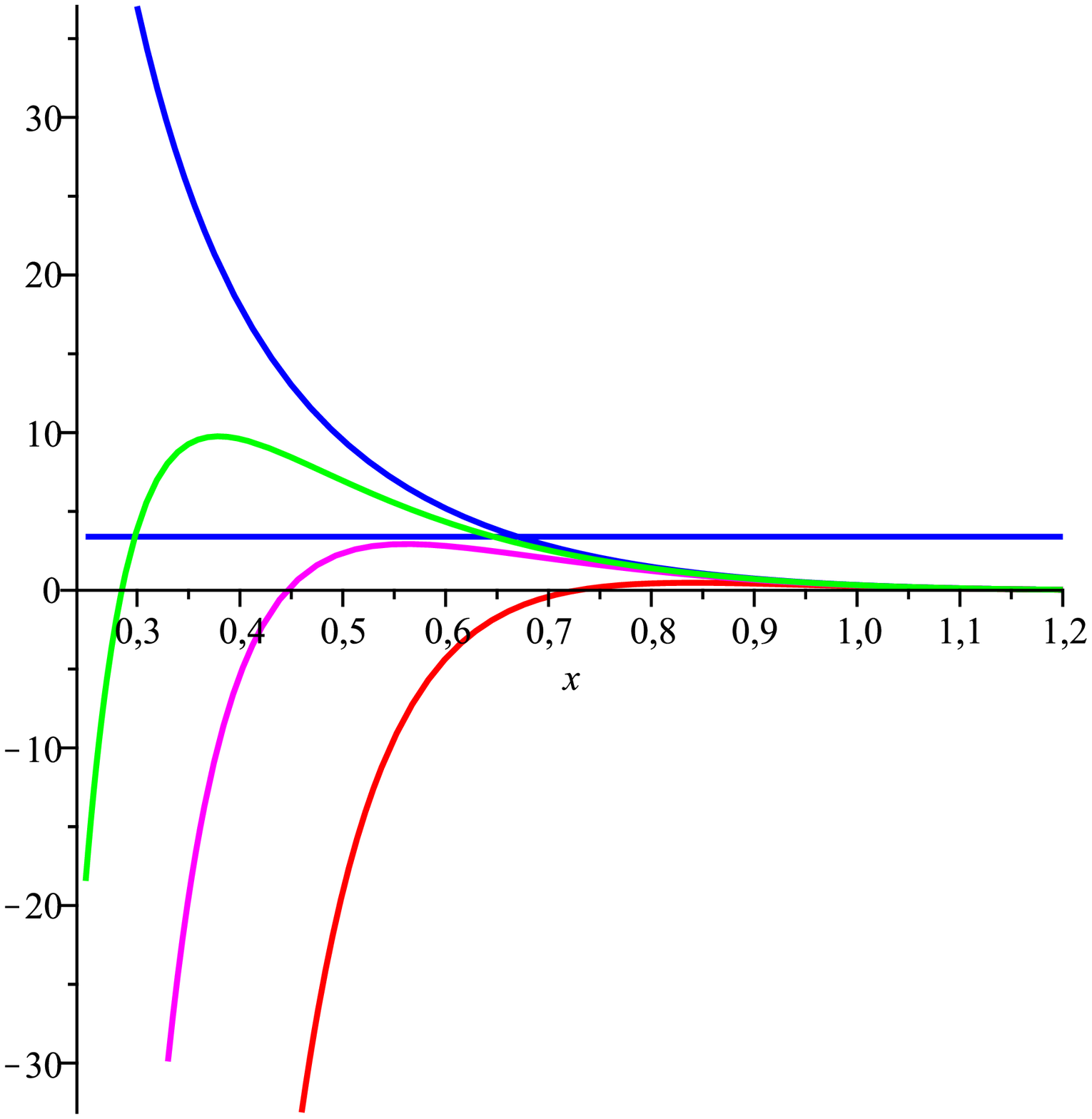} $\,\,\,\,\,\,\,$
B.
\includegraphics[height=6cm]{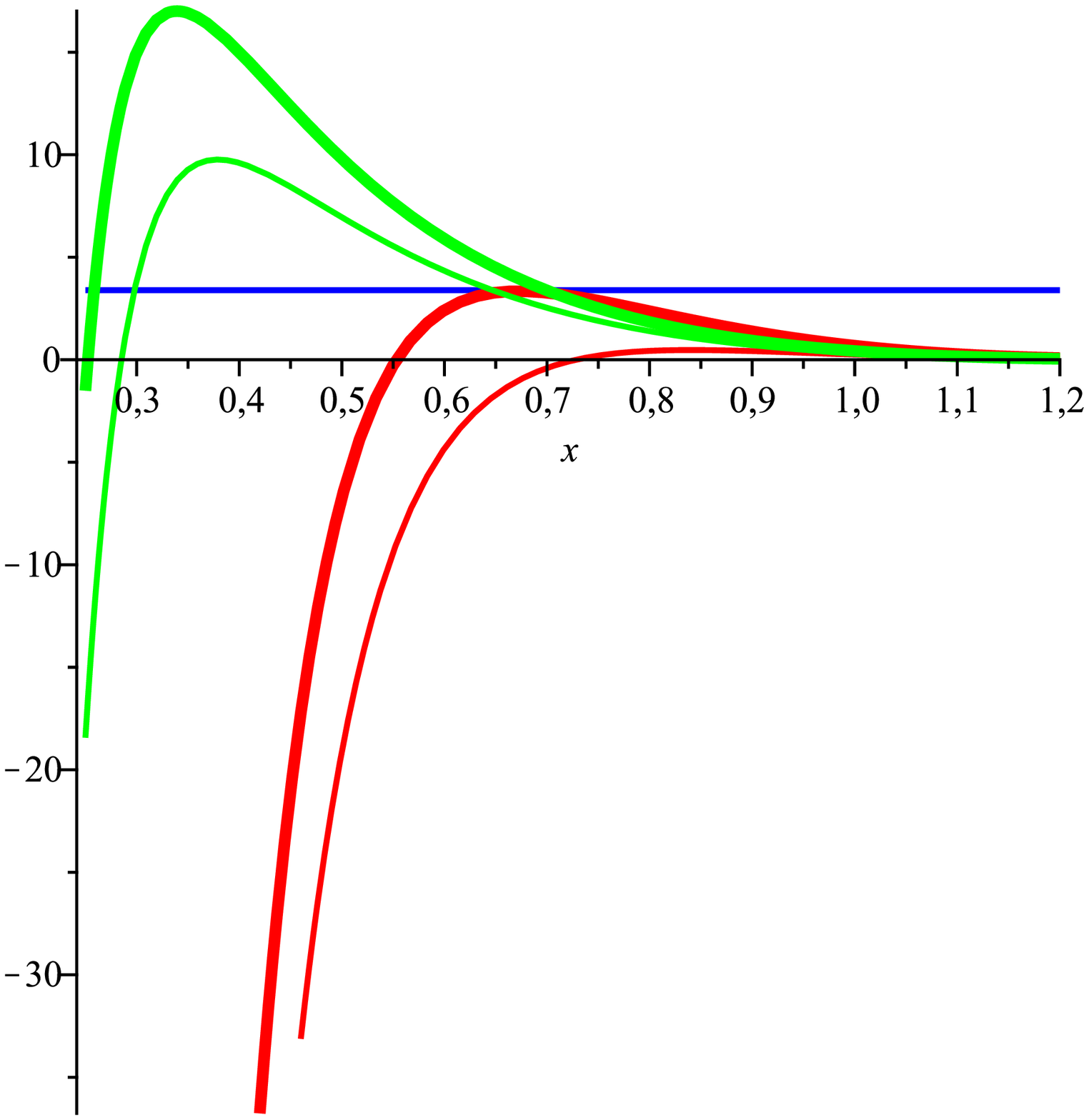}
\end{center}
\caption{
A. The left hand side of  equation
(3.2)
as function of $x$
   for the fixed value of $\bar{M}/a^2=1$ and different values of $\bar{Q}^2/a^4$.
   The blue line corresponds to $\bar{Q}=0$ and the straight blue line represents the
   right hand side of equation
   (3.2), i.e. $16/3\pi\kappa$ (for $\kappa=1/2$).
    The green line corresponds to
    $\bar{Q}^2/a^4<{\cal Q}_{cr}^2$. The magenta line represents the left hand side
    of equation
   (3.2) for
    $\bar{Q}^2/a^4={\cal Q}_{cr}^2$. The red line represents the left hand side  of equation
   (3.2) for
    $\bar{Q}^2/a^4>{\cal Q}_{cr}^2$ B.
     The thick green line corresponds to shock wave with higher value of $\bar{M}/a^2>1$
     and the same value of $\bar{Q}^2/a^4<{\cal Q}_{cr}^2$ as the fat green line. The thick red line
     corresponds to shock wave with higher value of $\bar{M}/a^2>1$
     and the same value of $\bar{Q}^2/a^4>{\cal Q}_{cr}^2$ as the fat red line.}
\label{fig:rho-AdS-0}
\end{figure}

In the case of $AdS_5$ background the shape function $F(\rho)$ is
\bea F_{5,AdS}(\rho)&=&\frac{3\sqrt{2}\pi\bar M}{a}\left( \frac{4a^4+12a^2\rho^2+\rho^4}{4a\rho (2a^2-\rho^2)}
-\sqrt{2}\frac{2a^2+\rho^2}{2a^2-\rho^2}\right)\\ &-&\frac{5\sqrt{2}\pi \bar Q^2}{a^3}\left(
\frac{16a^8-160a^6\rho^2-360a^4\rho^4-40a^2\rho^6+\rho^8}{256a^3\rho^3(2a^2-\rho^2)}+
\frac{\sqrt{2}}{2}\frac{2a^2+\rho^2}{2a^2-\rho^2} \right) \nonumber\eea
And the equation on the radius of trapped surface (\ref{rts-ads}) takes
the form (here we introduce $x=\rho_0/a$)

\be
\frac{(2-x^2)^3}{x^2(2+x^2)}\left(\frac{\bar{M}}{a^2}
-\frac{\bar{Q}^2}{a^4}
\frac{5}{64}\frac{(2-x^2)^2}{x^2}\right)=\frac{16}{3\pi\kappa}, \label{AdS-roots}
\ee
We solve this equation graphically, see  Fig.\ref{fig:rho-AdS-0}.  We assume
that value $\bar{M}/a^2$ is fixed,
 we  variate parameter
$\bar{Q}/a^2$ and analyze how the left hand side changes.
On the Fig.\ref{fig:rho-AdS-0} the blue straight line represents  the right
 hand side of (\ref{AdS-roots})
and another curves  represent the
left hand side of (\ref{AdS-roots}) for various values of the
 rescaled  charge $\bar{Q}^2/a^4$ and energy $\bar{M}/a^2$.

When $\bar{Q}=0$  equation (\ref{AdS-roots}) has only one  root
(intersection of two blue  lines in Fig.\ref{fig:rho-AdS-0}.A).
A  small  charge  leads to appearance of one new root
of equation
(\ref{AdS-roots}). This root is very close to zero  and the initial "chargeless"
 root moves to the left.
These two roots correspond to
two intersections  of the blue straight line and the green line in Fig.\ref{fig:rho-AdS-0}.A).
 Further increasing of the charge
makes for increasing  the smallest root and decreasing the biggest one.
At the critical value of
$\bar{Q}^2/a^4={\cal Q}_{cr}^2$
, ${\cal Q}_{cr}^2\simeq 0.5$,
 two roots coincide (the magenta line in Fig.\ref{fig:rho-AdS-0}.A) and
 we have only one root of equation
 (\ref{AdS-roots}). For  $\bar{Q}^2/a^4>{\cal Q}_{cr}^2$ equation (\ref{AdS-roots})  has
  no physical solution for given $\bar{M}/a^2$ and
trapped surface can not be formed. Fig.\ref{fig:rho-AdS-0}.B shows that by increasing
$\bar{M}/a^2$ we also increase ${\cal Q}_{cr}^2$.

Suppose that $\bar{Q}^2/a^4<{\cal Q}_{cr}^2$. The area of the trapped surface
is given by (for detailed derivation of similar formula in $dS$ see \cite{ABG})
 \bea
 {\cal A}_{AdS_5}&=&2\cdot Vol
S^{2}\int\limits_0^{\rho_0}
\frac{2\sqrt{2}\rho^{2}}{\left(1-\frac{\rho^2}{2a^2}
\right)^{3}}d\rho= \label{AadS5} \\ &=&4\sqrt{2}a^3\pi\left(\frac{2\rho_0
a(2a^2+\rho_0^2)}{(2a^2-\rho_0^2)^2}-\sqrt{2} \mbox{arctanh}
(\frac{\sqrt{2}\rho_0}{2a})\right). \nonumber
\eea
To estimate the area of the trapped surface we use the
biggest root of equation (\ref{AdS-roots}) because in the neutral limit this root tends to the "chargeless" one.

Firstly we assume that $\bar{M}<<a^2$. In this case a solution to
(\ref{AdS-roots})  is given by

\be
x^2\approx\frac{3\pi}{8}\frac{\bar{M}}{a^2}\left(2\kappa
-\frac{5\bar{Q}^2}{6\pi \bar{M}^2}
\right). \label{AdS-roots-apr}
\ee
For small $x=\rho/a$ the area is
\be {\cal A}_{AdS_5}\approx\frac{16\sqrt{2}\pi}{3}a^3x^3+{\cal O}(x^5)
\ee
and we get
\be
{\cal A}_{AdS_5}\approx\frac{16\sqrt{2}\pi a^3}{3}\left(\frac{3\pi}{8}\frac{\bar{M}}{a^2}\left(2\kappa
-\frac{5\bar{Q}^2}{6 \pi \bar{M}^2}
\right)\right)^{3/2}
\ee

Secondly we consider the case $\bar{M}>>a^2$. In this regime a
solution to (\ref{AdS-roots})  is given by $x=\sqrt{2}-y$, where
$y$ is small and satisfies the following equation

\be (y^3+\frac{3}{2\sqrt{2}}y^4+\frac{7}{8}y^5+...)\left(1
-\frac{\bar{Q}^2}{\bar{M}a^2}
\frac{5}{16}(y^2+...)\right)=\frac{4\sqrt{2}}{3\pi\kappa}\frac{a^2}{\bar{M}}.
\label{HE-roots} \ee Denoting the RHS of (\ref{HE-roots}) by $
\alpha^3\equiv\frac{4\sqrt{2}}{3\pi}\frac{a^2}{\kappa \bar{M}}$ we
find a perturbative solution \be y=\alpha
+p_1\alpha^2+p_2\alpha^3+... \ee where the coefficients are given
by \bea
p_1&=&-\frac{\sqrt{2}}{4}\\
p_2&=&\frac{1}{12}+\frac{\bar{Q}^2}{\bar{M}a^2}
\frac{5}{16\cdot3}
\eea
 The area of the trapped surface is given by
\bea
{\cal A}_{AdS_5}&\approx &4\sqrt{2} \pi a^3\left(\frac{\sqrt{2}}{y^2}-\frac{1}{y}+...\right)
\nonumber\\
&=&
4\pi a^3(\frac{3\pi\kappa}{2}\frac{\bar{M}}{a^2})^{2/3}\left( 1-\frac{1}{24}(1+\frac{5\bar{Q}^2}{\bar{M}a^2}
)\,(\frac{4\sqrt{2}}{3\kappa\pi}\frac{a^2}{\bar{M}})^{2/3}+...\right)
\eea
and we see that the first term reproduces the Gubser, Pufu and Yarom answer
\cite{Gubser08} and the charge decreases the area of the trapped surface.

\section{Solution to Trapped Surface Equation in $dS_4$}

 \subsection{Critical charges}

The profile of the charged shock wave in $dS_4$ in the chordal coordinate is given by

 \be
     \label{final-ds4}
 F_{4,dS}(\bar{M},\bar{Q}^2,q)=8\bar{M}\left[-1 +\frac{1-2q}{2}\ln\left(\frac{1-q}{q}\right)\right]-
 \bar{Q}^2\frac{3 \pi }{4a}\frac{1 -8q+8q^2}{\sqrt{q(1-q)}}
 \ee
 To visualize the solutions of equation (\ref{eq-q-dS})  for the shape function
 (\ref{final-ds4}) one can use the same method as in Sect. 3.
Let us  introduce the function
  \be
\label{eq-q-dS-m}
{\cal F}_{4,dS}(a,\bar{M},\bar{Q}^2,q)\equiv F'_{4,dS}(a,\bar{M},\bar{Q}^2,q)+
\frac{2F_{4,dS}(a,\bar{M},\bar{Q}^2,q)}{1-2q}+
 \frac{2a}{\sqrt{q(1-q)}}.
\ee
It can be presented as
\be
\label{d-n}
{\cal F}_{4,dS}(a,\bar{M},\bar{Q}^2,q)=\frac{{\cal N}_{4,dS}(a,\bar{M},\bar{Q}^2,q)}
{{\cal D}_{4,dS}(a,q)},
\ee
where
\be
{\cal N}_{4,dS}(a,\bar{M},\bar{Q}^2,q)={\cal N}_{4,dS}(a,\bar{M},q))+
 3 \pi \frac{\bar{Q}^2 }{a^2},
 \ee
and
\bea
{\cal N}_{4,dS}(a,\bar{M},q)=32  q^3 -48  q^2 +16  q -32 \frac{\bar{M}}{a}
 \sqrt{q (1-q)}.
\eea
To solve equation
\be
\label{eq-q-ds}
{\cal F}_{4,dS}(a,\bar{M},\bar{Q}^2,q)=0\ee
one can
 draw the function $-{\cal N}_{4,dS}(\bar{M}/a,q)$
 and see where it can be  equal to $3 \bar{Q}^2/a^2 \pi $.

\begin{figure}[h]
    \begin{center}
A.\includegraphics[width=6cm]{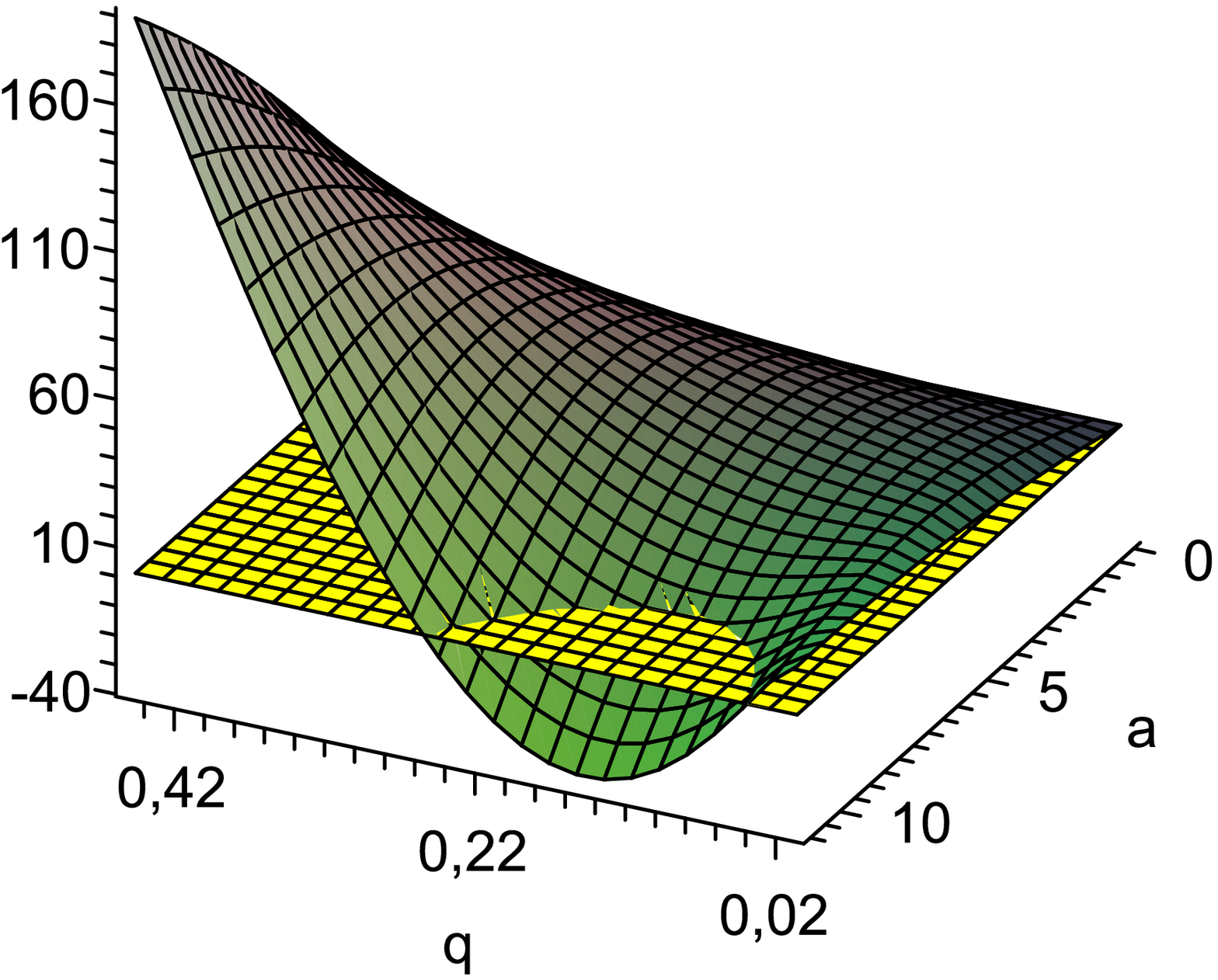}
$\,$B
\includegraphics[width=6cm]{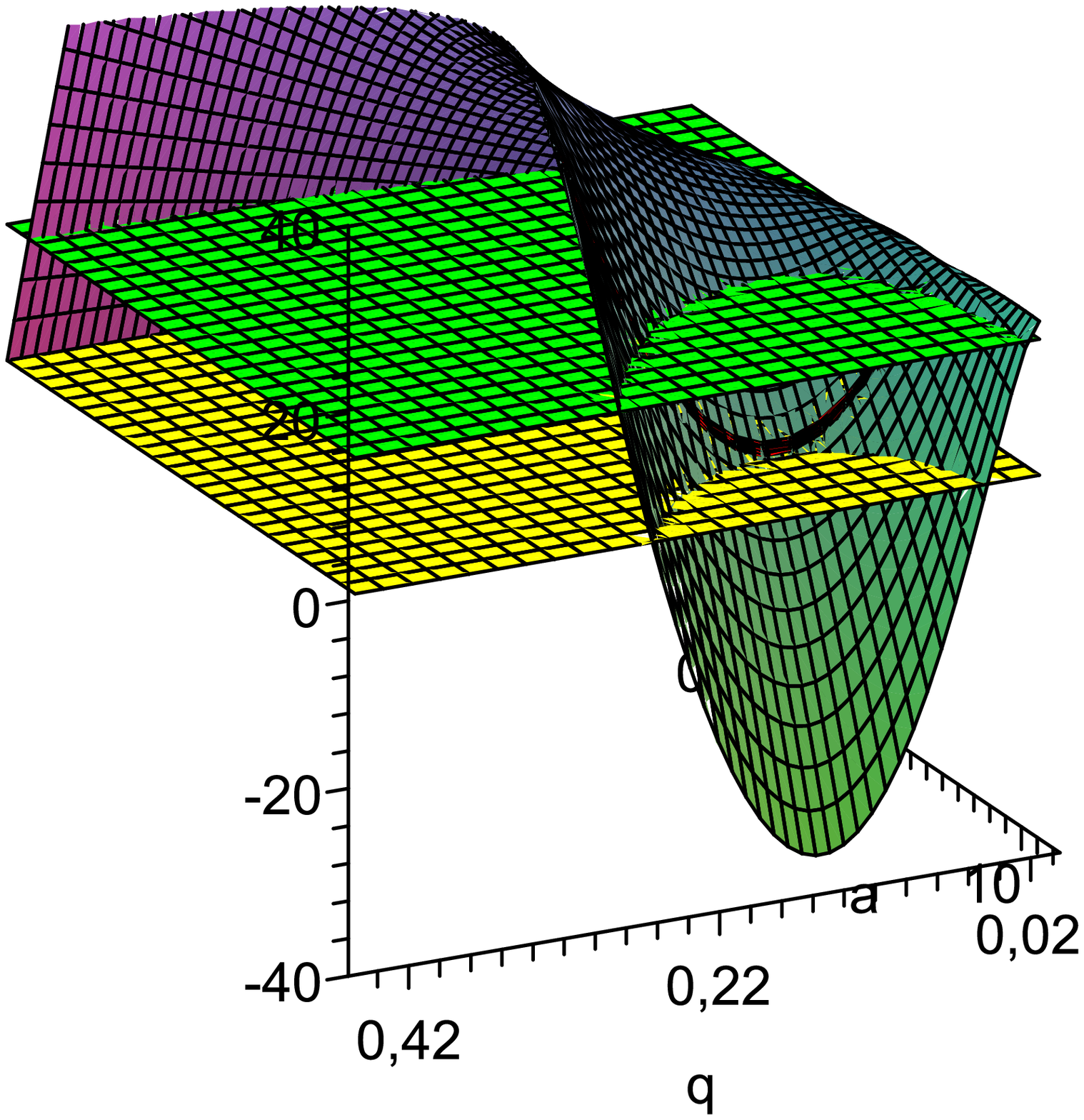}
\end{center}
   \caption{A. The multicolored  surface represents
   $-{\cal N}_{4,dS}(a,\bar{M},q)$ as function of two variables $q$
   and $a$; $\bar{M}=1$. The intersection with the yellow plane
   shows the existence of solutions of equation (4.2) for $Q=0$ for $a>a_{cr,0}$.
   For $0<a<a_{cr,Q}$ there is no intersection of the yellow plane with the
   multicolored  surface
   B.
   The intersection the multicolored  surface with
   the green plane shows
   the existence of solutions of equation (4.2) for given $\bar{Q}$ for $a<a_{cr,Q}$.
   We see that the critical value of the
   gravitational radius  for two colliding   chargeless shock waves
   is smaller as compare with the critical  radius for  charged shock waves
   with the same energy. }
\label{picture-crit-dS-3d}\end{figure}

The shape of $-{\cal N}_{4,dS}(\bar{M}/a,q)$  for a fixed value of $\bar{M}=1$
is presented in Fig.\ref{picture-crit-dS-3d}.
In Figure \ref{picture-crit-dS-3d}.A. we can see results obtained in \cite{ABG},
namely the existence of a critical value of $a=a_{cr,0}(\bar{M})$, below which the trapped
surface formation is impossible. Indeed, for $a$ small enough  values of
this function are positive  and equation (\ref{eq-q-dS-m}) has no solution.
The charge makes this critical value of $a=a_{cr,Q^2}(\bar{M})$ for the same
energy less then the corresponding
$a_{cr,0}(\bar{M})$ for  the chargeless case.
(see Fig.\ref{picture-crit-dS-3d}.B).

\begin{figure}[h]
    \begin{center}
A.\includegraphics[height=5cm]{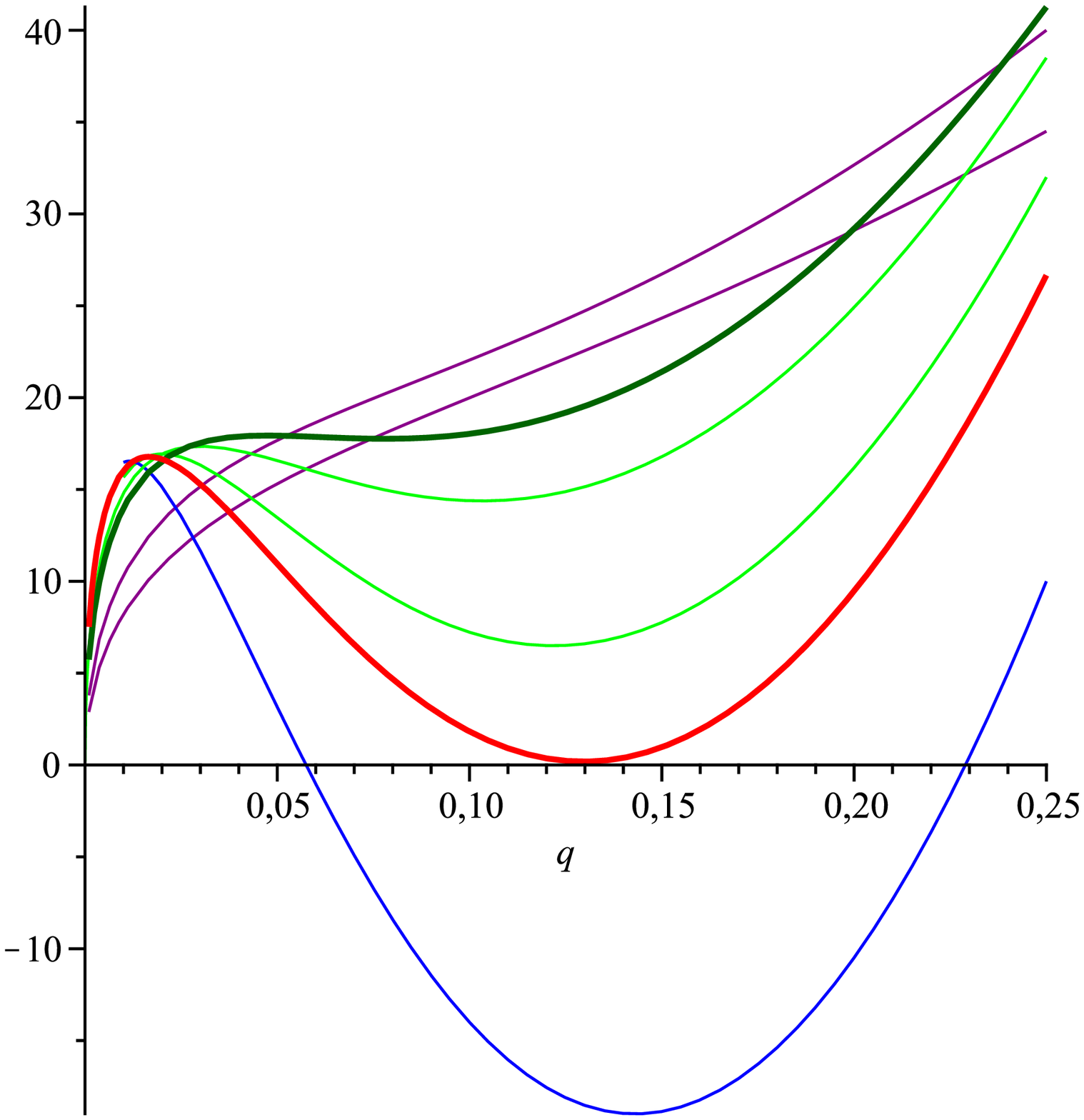}
 $\,\,\,\,\,\,\,$B.\includegraphics[height=5cm]{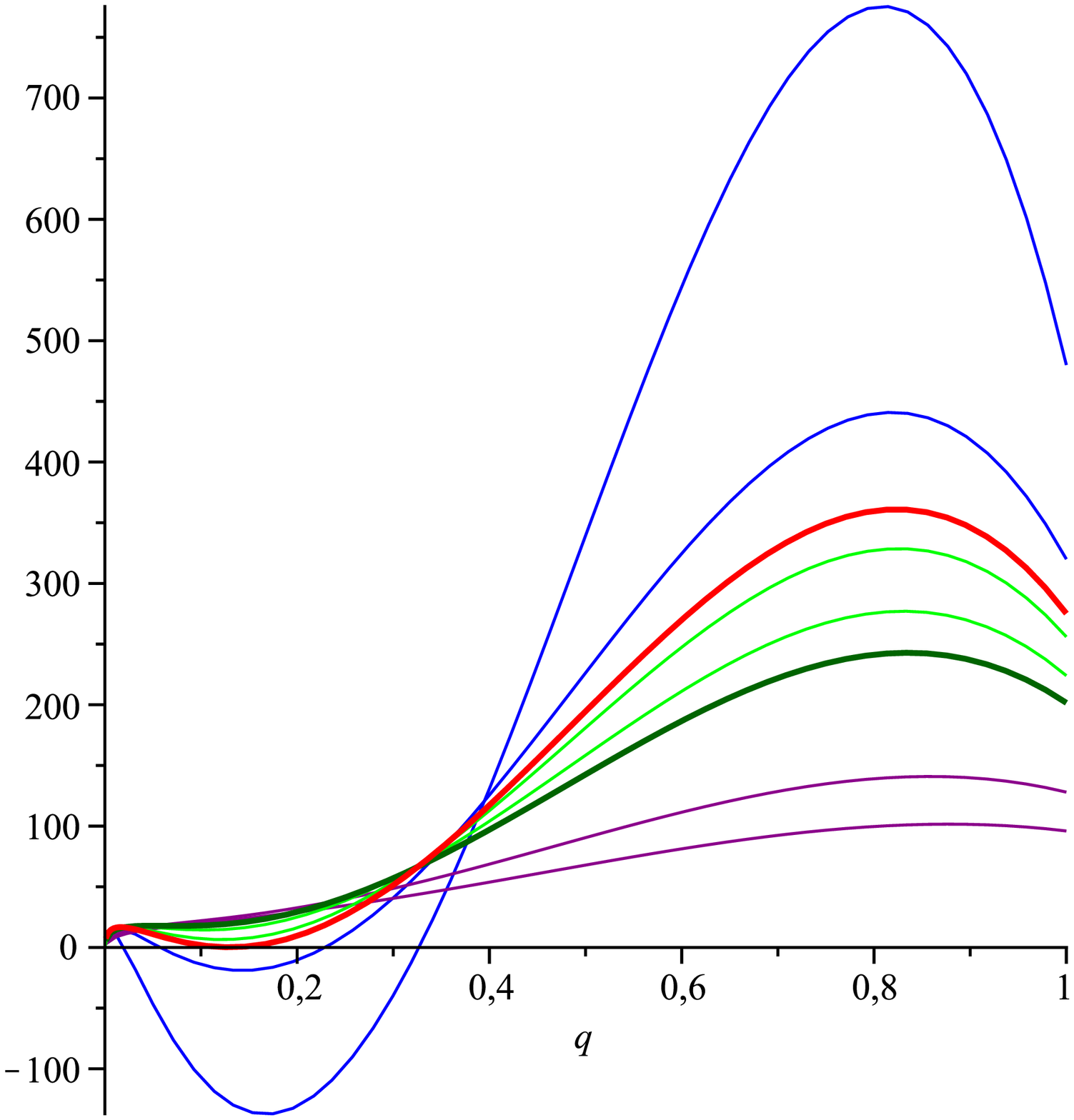}
\end{center}
   \caption{A. The shape of functions $-{\cal N}_{4,dS}(a,\bar{M},q)$ as
   functions of $q$
   for different fixed values of $a$, $\bar{M}=1$. The red line presents the shape for
    $a$ equal to
   the critical
   value $a_{cr,0}$. The blue line presents the shape for  $a>a_{cr,0}$,
   magenta and green lines correspond to $a<a_{cr,0}$;
   the dark green line and  magenta lines have not nontrivial local minimum.
    B. Here is shown that all curves in the region $0<q<1$ have  maxima  that
    demonstrates the existence of the critical values of the charge
    above which there is no trapped surface formation. }
\label{picture-crit-dS}\end{figure}

In Fig.\ref{picture-crit-dS}
the shape of  function $-{\cal N}_{4,dS}(a,\bar{M},q))$ as a
functions of $q$
for different fixed values of $a$ and $\bar{M}=1$ is presented.

Fig.\ref{picture-crit-dS}.A. shows that  in the case of  $a<a_{cr,0}(\bar{M})$, i.e.
in the case when chargeless version of equation (\ref{final-ds4}) has not solution,
the presence of the charge drastically change the situation.
 Already a small charge produces a nonzero  solution of the equation (\ref{eq-q-ds}).
By increasing the charge we reach a domain where charge effect dominates,
i.e. the second term in the RHS of (\ref{final-ds4})
dominates and equation  (\ref{eq-q-ds}) has no  solutions.

Fig.\ref{picture-crit-dS}.B. shows that for $a>a_{cr,0}(\bar{M})$ there is
the maximum value of
the function $-{\cal N}_{4,dS}(a,\bar{M},q)$. This maximum for different $a$
defines critical values of $\bar{Q}^2$,
${\cal Q}_{max}^2={\cal Q}_{max}^2(a,\bar{M})$,
above which there is no formation of the trapped surface.
 Similar maximum have curves representing the case of
 $a<a_{cr,0}$ (the green and magenta lines
 in
Fig.\ref{picture-crit-dS}.B).

\subsection{Area of the trapped surface in  $dS_4$}

In $dS_4$ background the shape is
\be F_{4,dS}(\rho)=4\bar M \left( -2+\frac{2a^2-\rho^2}{2a^2+\rho^2}\ln \frac{2a^2}{\rho^2} \right)-
\frac{3\sqrt{2}\pi \bar Q^2}{8a^2} \frac{4a^4-12a^2\rho^2+\rho^4}{\rho(2a^2+\rho^2)},\ee

and equation (\ref{rts-ds}) has the form:
\be
\frac{(2+x^2)^2}{x(2-x^2)}\left( \frac{\bar{M}}{a}-\frac{\bar{Q}^2}{2a^2}\frac{3\sqrt{2}\pi}{32} \frac{2+x^2}{x}
\right)=\frac{\sqrt{2}}{2\kappa}.
 \label{dS-roots}
 \ee

For $\bar{Q}^2=0$  there are two possibilities.  Whether $\bar{M}>a/4$ and
 we have not a solution to the
equation on the radius of trapped surface, or  $\bar{M}\leq a/4$
and two solution to this equation exist (see \cite{ABG}). We take
the smaller one because the larger is localized near spatial infinity. The graphical solution of this equation is
presented in Fig.7. The LHS of (\ref{dS-roots}) for $\bar{M}>a/4$ is drawn by
the red line in Fig.7.A and for $\bar{M}<a/4$ by
the magenta line in Fig.7.B. The RHS of
(\ref{dS-roots}) is presented by the green straight lines.

 Let us consider graphically as well the influence of charge.
\begin{figure}[h]
    \begin{center}
A.\includegraphics[height=6cm]{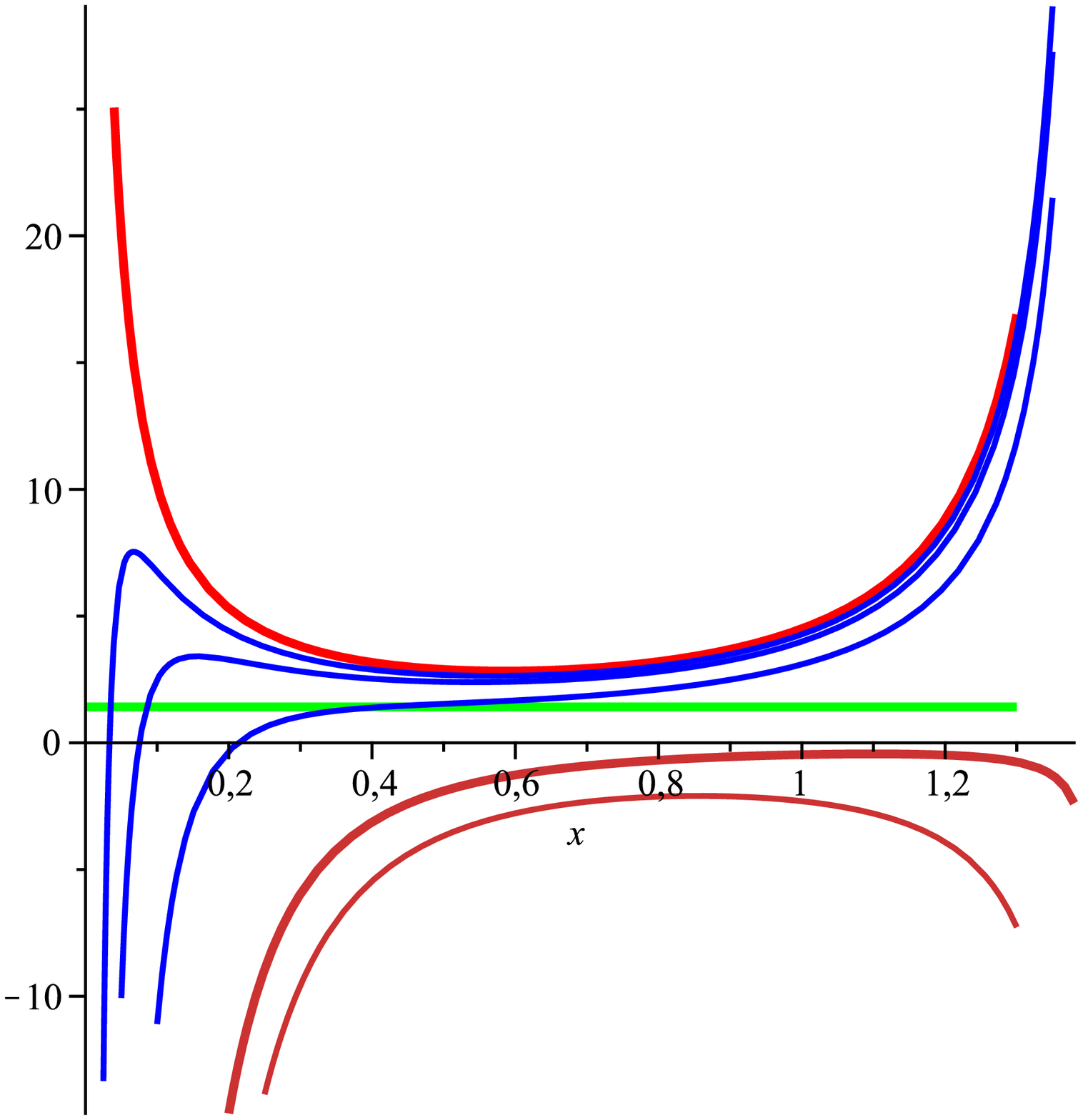}$\,\,\,\,\,\,$B.
\includegraphics[height=6cm]{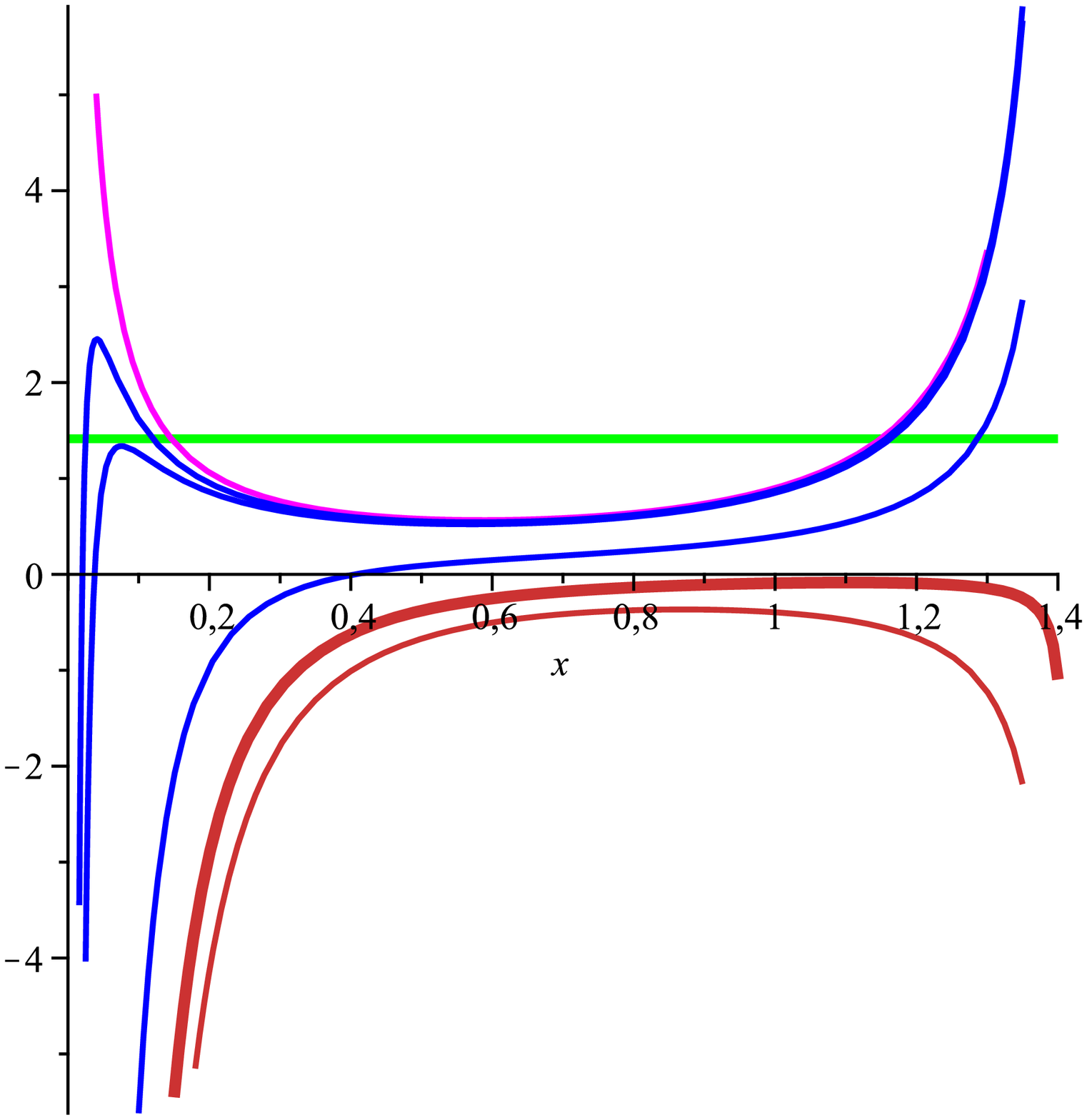}
 \caption{A. $a<a_{cr,0}$.
 The LHS
   of   eq. (4.2) for
    $\bar{M}/a=0.5>\mu_{cr,0}$
   and different  values of $\bar{Q}^2/a^2$: $\bar{Q}^2/a^2=0$ (red line),
   $0<\bar{Q}^2/a^2<{\cal Q}^2_{cr}\approx 0.94$ (blue lines),  and $\bar{Q}^2/a^2>{\cal Q}^2_{cr}$ (orange lines)
   The green straight line represents the value of the RHS of eq.(4.2). B. $a>a_{cr,0}$. LHS of   eq.(4.2) for
   $\bar{M}/a=0.1<\mu_{cr,0}$
   and different  values of $\bar{Q}^2/a^2$ from  $\bar{Q}^2/a^2=0$ (magenta line),
   $0<\bar{Q}^2/a^2<{\cal Q}^2_{cr}\approx 0.41$ (blue lines),  and $\bar{Q}^2/a^2>{\cal Q}^2_{cr}$ (orange lines).
   The green line represents the value of the RHS of eq.(4.2)}\end{center}
   \label{fig:crit-dS}
\end{figure}

In Fig.7 different lines correspond to the left hand side of
(\ref{dS-roots}) for various values of charge $\bar{Q}$.

Let us consider first $a<a_{cr,0}$, Fig.7.A. Note that
 the function representing the case $\bar{Q}=0$ has a positive singularity at $x=0$
 (the red curve).
If $\bar{Q}\neq 0$ and is arbitrary small then the sign of the singularity at $x=0$
changes and this fact leads to appearance of
one root. Growth of the $\bar{Q}$ makes for growth of the root
(the set of blue curves in Fig.7.A; lower curve corresponds to
the bigger value of $\bar{Q}^2$).

Further increasing of $\bar{Q}^2$ causes the change of the sign of the second
singularity at $x=\sqrt{2}$
and we have no more intersection points between the green line and the
lines representing the left hand side of (\ref{dS-roots}). The trapped surface cannot
be formed  as in the neutral case.

Slightly different behavior of roots can be observed in the case,
when the trapped
surface exists already for $\bar{Q}=0$.

When  equation (\ref{dS-roots}) for $\bar{Q}=0$ has two roots
 (intersection of the green and magenta lines), i.e.  $a>a_{cr,0}$ (Fig.7.B),
arbitrary small increasing of the charge  leads to appearance of one new roots of
 (\ref{dS-roots}) and to shift of the existent roots. Both of two new small roots
are smaller than initial (intersection of green and the upper blue line).
The third root near $x=\sqrt{2}$ seems nonphysical
because it is localized near spatial infinity. At some critical value of $\bar{Q}$
(the second blue line) only one small root remains, and after that we have no
physical solution to the equation for
trapped surface radius. And if $\bar{Q}$ becomes much larger (brown curves with two negative singularities)
there is no solution, even nonphysical.

We now assume that $\bar{Q}^2\leq \bar{Q}^2_{cr}$ and $\bar{M}\ll
a$. In this case solution to (\ref{dS-roots}) is very close to
$x_0=\sqrt{2}\frac{\bar{M}}{a}$ \be \rho_0(\bar{M}/a,
\bar{Q}^2/a^2) \approx
2\sqrt{2}\,\kappa\bar{M}(1-\sqrt{2}\frac{3\pi}{64}
\frac{\bar{Q}^2}{\kappa\bar{M}^2}) \ee i.e. the presence of the
charge decreases the radius of the formed trapped surface.

The area of the trapped surface
is given by
\be {\cal A}_{dS_4}=2\cdot Vol
S^{1}\int\limits_0^{\rho_0}
\frac{2\rho}{\left(1+\frac{\rho^2}{2a^2}
\right)^{2}}d\rho=8\pi
\frac{a^2\rho_0^2}{2a^2+\rho_0^2}\ee

For $\bar{Q}^2\leq Q^2_{cr}$ and $\bar{M}\ll a$ \be {\cal
A}_{dS_4}\approx 4\pi \rho_0^2(1- \frac{\rho_0^2}{2a^2})\approx
32\pi\kappa^2 \bar{M}^2(1-\frac{4\kappa^2\bar{M}^2}{a^2}-
\frac{3\pi}{64}\frac{\bar{Q}^2}{\kappa\bar{M}^2}))\ee

\section{Conclusion}

 We have presented   the charged version of Aichelburg-Sexl
metric in the $(A)dS$ case.
 We have calculated  the area of the trapped
surface produced in a head-on collision of two charged shock waves in $(A)dS$ background.
We observe that the physical picture in $AdS$ is similar to the flat case. Namely, there is
 a critical value of charge
above which no marginally trapped surface on the past light cone is formed in the head-on collision. The phenomenon is analogous to the
the critical behavior found in flat space \cite{YosMann}.  The value of critical charge
in the $AdS$ case depends  both on
the collision energy and the value of the cosmological constant.

It would be interesting to study non-head-on collisions of charged particles
in $AdS$ background and, in particular, to study an influence of charges on
a possible elongation of  the shape of the  trapped surface  in
spherical coordinate.

We have found new interesting phenomena in the collision of shock
waves in $dS$ background. As has been already noticed in
\cite{ABG} there is a critical value of a cosmological radius
below which there is no trapped surfaces formation. Non
zero charge changes situation and formation of trapped surface
becomes possible even if in the neutral case for given values of $\bar M$ and $a$ the ratio $\bar M/a$ is over critical point. However a large
enough charge stops a formation of the trapped surface. It would
be also interesting to study non-head-on collisions of charged
particles in $dS$ background and to find how a non-zero impact
parameter influences on the trapped surface formation in the
collision of charged particles.

\section*{Acknowledgments}

I.A. is grateful to I.Volovich for fruitful discussions.
I.A.  and A.B. are supported in part by RFBR grant 08-01-00798, I.A. is supported also
in part by grant
NS-795.2008.1. L.J. acknowledges the support of the Centre for Theoretical Cosmology, in Cambridge.

\newpage
\appendix
\section{Appendix}
\subsection{Standard form of Lemma}
{\bf Lemma I.} For an integrable function $f$ takes place the identity
\be
    \label{lim}
    \lim _{v\to 1}\gamma f\left(\gamma^2 (Y_{0}+v Y_{1})^2\right)=\delta(Y_{0}+ Y_{1})
    \int f(x^2)dx
\ee
This one  might be found in  \cite{HT}.
\subsection{Modified  form of Lemma}

As has been mentioned in the text  in the dS case we have to deal with
 expressions which require a regularization, in particular with  expression
(\ref{lsin}).
Let us first consider a  simple example and prove  the following

{\bf Lemma II. }
{\it In the sense of distributions one has}
\bea
\label{lim-sin-l}
\lim _{\gamma\to \infty}\left(\frac{\gamma }{Z^2-\gamma^2Y^2}\right)_{reg}f(\gamma^2 Y^2)
 =\delta(Y)\int \left(\frac{1}{Z^2-x^2}\right)_{reg}f( x^2)dx
\eea
{\it here we  use the following regularization}
\bea
\label{reg}
&\,&(\left(\frac{\gamma }{Z^2-\gamma^2Y^2}\right)_{reg}f(\gamma^2 Y^2),g)
\label{def}\\
&=&\int_{|Z-x|<1}\frac{1}{2Z} \frac{f(x^2)g(\frac{x}{\gamma})-f(z^2)g(\frac{z}{\gamma})}{Z-x}dx+
\int_{|Z-x|>1}\frac{1}{2Z} \frac{f(x^2)g(\frac{x}{\gamma})}{Z-x}dx\nonumber\\
&+&\int_{|Z+x|<1}\frac{1}{2Z} \frac{f(x^2)g(\frac{x}{\gamma})-f(z^2)g(-\frac{z}{\gamma})}{Z+x}dx+
\int_{|Z+x|>1}\frac{1}{2Z} \frac{f(x^2)g(\frac{x}{\gamma})}{Z+x}dx\nonumber\eea

{\bf Remark 1} To write  regularization (\ref{reg}) we assume:\\
i) the following
relation
\be \label{under}\int \left(\frac{\gamma
}{Z^2-\gamma^2Y^2}\right)_{reg}f(\gamma^2 Y^2)g(Y)dy \equiv
\int
\left(\frac{1}{Z^2-x^2}\right)_{reg}f(x^2)g(\frac{x}{\gamma})dx
\ee
ii) the regularization in the RHS of (\ref{under})
 in the following sense
\bea
\label{def-1}
\int
\left(\frac{1}{Z^2-x^2}\right)_{reg}f(x^2)dx
=\left(\int \frac{f(x^2)}{2Z} (\frac{1}{Z-x})_{reg}dx
+
\int
\frac{f(x^2)}{2Z}( \frac{1}{Z+x})_{reg}dx\right)
\eea
iii) the natural regularization  for two terms in (\ref{def-1})
\bea
\label{def-2}\int (\frac{1}{Z-x})_{reg}\, f(x^2)dx
=\left(\int_{|Z-x|<1}\frac{f(x^2)-f(z^2)}{Z-x}dx+
\int_{|Z-x|>1} \frac{f(x^2)}{Z-x}dx\right)\eea
\bea
\int
( \frac{1}{Z+x})_{reg}\,f(x^2)dx
+\left(\int_{|Z+x|<1} \frac{f(x^2)-f(Z^2)}{Z+x}dx+
\int_{|Z+x|>1} \frac{f(x^2)}{Z+x}dx\right).\label{def-3a}
\eea

{\bf Remark 2} For simplicity we have made a cut-off in (\ref{def-3a}) and
(\ref{def-2}) as well in (\ref{reg}) at $1$,
but one can make it at an arbitrary {\it dimensional}
parameter, say $\epsilon$.

{\bf Proof} $\,$ Taking  the limit $\gamma \to \infty$ in the RHS of (\ref{def})
 we get
\bea
&\,&\lim _{\gamma \to \infty}(\left(\frac{\gamma }{Z^2-\gamma^2Y^2}\right)_{reg}f(\gamma^2 Y^2),g)\\\nonumber
&=&\lim _{\gamma \to \infty}\left(\int_{|Z-x|<1}\frac{1}{2Z} \frac{f(x^2)g(\frac{x}{\gamma})-f(z^2)g(\frac{z}{\gamma})}{Z-x}dx+
\int_{|Z-x|>1}\frac{1}{2Z} \frac{f(x^2)g(\frac{x}{\gamma})}{Z-x}dx\right)\\\nonumber
&+&\lim _{\gamma \to \infty}\left(\int_{|Z+x|<1}\frac{1}{2Z} \frac{f(x^2)g(\frac{x}{\gamma})-f(z^2)g(-\frac{z}{\gamma})}{Z+x}dx+
\int_{|Z+x|>1}\frac{1}{2Z} \frac{f(x^2)g(\frac{x}{\gamma})}{Z+x}dx\right)\\\nonumber
&=&\int_{|Z-x|<1}\frac{1}{2Z} \frac{f(x^2)g(0)-f(z^2)g(0)}{Z-x}dx+
\int_{|Z-x|>1}\frac{1}{2Z} \frac{f(x^2)g(0)}{Z-x}dx\\\nonumber
&+&\int_{|Z+x|<1}\frac{1}{2Z} \frac{f(x^2)g(\frac{x}{\gamma})-f(z^2)g(0)}{Z+x}dx+
\int_{|Z+x|>1}\frac{1}{2Z} \frac{f(x^2)g(0)}{Z+x}dx\\\nonumber
&=&g(0)\left(\int_{|Z-x|<1}\frac{1}{2Z} \frac{f(x^2)-f(z^2)}{Z-x}dx+
\int_{|Z-x|>1}\frac{1}{2Z} \frac{f(x^2)}{Z-x}dx\right)\\\nonumber
&+&g(0)\left(\int_{|Z+x|<1}\frac{1}{2Z} \frac{f(x^2)-f(Z^2)}{Z+x}dx+
\int_{|Z+x|>1}\frac{1}{2Z} \frac{f(x^2)}{Z+x}dx\right)\\\nonumber
&=&g(0)\left(\int \frac{f(x^2)}{2Z} (\frac{1}{Z-x})_{reg}dx+\int
\frac{f(x^2)}{2Z}( \frac{1}{Z+x})_{reg}dx\right)\\\nonumber
\eea
that is nothing but
\be
g(0)\int f(x^2)(\frac{1}{Z^2-x^2})_{reg}dx
\ee
that proves the Lemma II.

Let us prove the Lemma that is suitable for dS case.

{\bf Lemma III} {\it In the sense of distributions one has}
\bea
\label{lim-sin-l}
\lim _{\gamma\to \infty}\left(\frac{\gamma }{(Z^2-\gamma^2Y^2)^2}\right)_{reg}f(\gamma^2 Y^2)
 =\delta(Y)\int \left(\frac{1}{(Z^2-x^2)^2}\right)_{reg}f( x^2)dx
\eea
{\it We  use the following regularization}

\bea
&\,&(\left(\frac{\gamma }{(Z^2-\gamma^2Y^2)^2}\right)_{reg}f(\gamma^2 Y^2),g)
\label{deff}\\\nonumber
&=&\int_{|Z-x|<1}\frac{1}{2(Z^2+x^2)} \frac{f(x^2)g(\frac{x}{\gamma})-f(Z^2)
g(\frac{Z}{\gamma})-
\frac{\partial}{\partial Z}\left(f(Z^2)g(\frac{Z}{\gamma})
\right)(Z-x)}{(Z-x)^2}dx\\\nonumber&+&
\int_{|Z-x|>1}\frac{1}{2(Z^2+x^2)} \frac{f(x^2)g(\frac{x}{\gamma})}{(Z-x)^2}\, dx
\\\nonumber
&+&\int_{|Z+x|<1}\frac{1}{2(Z^2+x^2)} \frac{f(x^2)g(\frac{x}{\gamma})-
f(Z^2)g(-\frac{Z}{\gamma})
-\frac{\partial}{\partial Z}\left(f(Z^2)g(-\frac{Z}{\gamma})\right)
(Z+x)}{(Z+x)^2}dx\\\nonumber
&+&
\int_{|Z+x|>1}\frac{1}{2(Z^2+x^2)} \frac{f(x^2)g(\frac{x}{\gamma})}{(Z+x)^2}
\,dx
\eea
{\bf Remark 3.} To write this regularization we are motivated by the following relation
\be
\int \left(\frac{\gamma }{(Z^2-\gamma^2Y^2)^2}\right)_{reg}f(\gamma^2 Y^2)g(Y)\,dy
\equiv\int \left(\frac{1}{(Z^2-x^2)^2}\right)_{reg}f(x^2)g(\frac{x}{\gamma})\, dx
\ee

{\bf Proof}
Using
\be
\frac{\partial}{\partial Z}\left(f(Z^2)g(\frac{Z}{\gamma})\right)=2Zf^\prime(Z^2)g(\frac{Z}{\gamma})
+\frac{1}{\gamma}f(Z^2)g^\prime(\frac{Z}{\gamma})
\ee
here
\be
f^\prime(x)=\frac{\partial}{\partial x}f(x)
\ee
and taking  the limit $\gamma \to \infty$ in the RHS of (\ref{deff}) we get
\bea
&\,&\lim _{\gamma \to \infty}(\left(\frac{\gamma }{(Z^2-\gamma^2Y^2)^2}f(\gamma^2 Y^2)\right)_{reg},g)\\
&=&\lim _{\gamma \to \infty}\left(\int_{|Z-x|<1}\frac{1}{2(Z^2+x^2)} \frac{f(x^2)g(\frac{x}{\gamma})
-f(Z^2)g(\frac{Z}{\gamma})-\frac{\partial}{\partial Z}\left(f(Z^2)g(\frac{Z}{\gamma})\right)(x-Z)}{(Z-x)^2}\,dx\right.\nonumber\\&+&
\left.\int_{|Z-x|>1}\frac{1}{2(Z^2+x^2)} \frac{f(x^2)g(\frac{x}{\gamma})}{(Z-x)^2}\,dx\right)\nonumber
\\
&+&\lim _{\gamma \to \infty}\left(\int_{|Z+x|<1}\frac{1}{2Z} \frac{f(x^2)g(\frac{x}{\gamma})-
f(Z^2)g(-\frac{Z}{\gamma})-\frac{\partial}{\partial Z}\left(f(Z^2)g(-\frac{Z}{\gamma})\right)(Z+x)}{Z+x}\,dx
\right.\nonumber\\&+&
\left.\int_{|Z+x|>1}\frac{1}{2(Z^2+x^2)} \frac{f(x^2)g(\frac{x}{\gamma})}{(Z+x)^2}\,dx\right)\nonumber\\
&=&g(0)\left(\int_{|Z-x|<1}\frac{1}{2(Z^2+x^2)} \frac{f(x^2)-f(Z^2)-\left(\frac{\partial}{\partial Z}f(Z^2)\right)(Z-x)}{(Z-x)^2}\,dx\right.\\
&+&
\left.\int_{|Z-x|>1}\frac{1}{2(Z^2+x^2)} \frac{f(x^2)}{(Z-x)^2}dx\right)\nonumber\\
&+&g(0)\left(\int_{|Z+x|<1}\frac{1}{2(Z^2+x^2)} \frac{f(x^2)-f(Z^2)-\left(\frac{\partial}{\partial Z}f(Z^2)\right)(Z+x)}{(Z+x)^2}\,dx
\right.\\
&+&
\left.\int_{|Z+x|>1}\frac{1}{2(Z^2+x^2)} \frac{f(x^2)}{(Z+x)^2}dx\right)\nonumber\\
&=&g(0)\left(\int \frac{f(x^2)}{2(Z^2+x^2)} (\frac{1}{(Z-x)^2})_{reg}\,dx+\int
\frac{f(x^2)}{2(Z^2+x^2)}( \frac{1}{(Z+x)^2})_{reg}dx\right)\nonumber\\
&\equiv&g(0)\int f(x^2)(\frac{1}{(Z^2-x^2)^2})_{reg}\,dx\eea
In the above calculations we use that the coefficient in front of
the derivative of $g$ goes to zero when
$\gamma\to\infty$.

{\bf Remark 4.} One can replace "1" in the domain of the integration in
(\ref{deff})
 by some dimensional parameter, say $\epsilon$ and the regularization
  prescription which we use in fact
 means what we deal with the principle values of integral and just remove
 $\epsilon $ neighborhood in integrals near points $x=\pm Z_D$. This prescription
 give D-dimensional answers presented in \cite{HT}.


\newpage
\begin{thebibliography}{99}

\bibitem{thooft_s}
 G.~'t~Hooft {\it  Gravitational collapse and particle physics},
 Proceedings: Proton-antiproton Collider Physics, Aachen,
1986,
 pp.~669--688;\\
 G.~'t Hooft,
  {\it Graviton Dominance in Ultrahigh-Energy Scattering},
  Phys.\ Lett.\  B {\bf 198} (1987) 61;
  %%CITATION = PHLTA,B198,61;%%
\\
 G.~'t Hooft,
  {\it On the factorization of universal poles in a theory of gravitating point
  particles},
  Nucl.\ Phys.\  B {\bf 304} (1988) 867.
  %%CITATION = NUPHA,B304,867;%%

\bibitem{Dray} T. Dray and G. 't Hooft, "The Gravitational Shock Wave of a Massless Particle,"
Nucl. Phys. B253 (1985) 173.

\bibitem{ACV}
D.~Amati, M.~Ciafaloni and G.~Veneziano,
{\it Superstring collisions at Planckian energies,}
  \PL   {\bf B197} (1987) 81.
  %%CITATION = PHLTA,B197,81;%%
\\
D.~Amati, M.~Ciafaloni and G.~Veneziano,
{\it Classical and quantum gravity effects from Planckian energy superstring
 collisions,}
Int.\ J.\ Mod.\ Phys.\  {\bf A3} (1988) 1615.
  %%CITATION = IMPAE,A3,1615;%%
\\
D.~Amati, M.~Ciafaloni and G.~Veneziano,
{\it Can space-time be probed below the string size?}
  \PL\  {\bf B216} (1989) 41.
  %%CITATION = PHLTA,B216,41;%%
\\
D.~Amati, M.~Ciafaloni and G.~Veneziano,
{\it Higher order gravitational deflection and soft bremsstrahlung in Planckian
energy superstring collisions,}
  \NP\  {\bf B347} (1990) 550.
  %%CITATION = NUPHA,B347,550;%%
\\
D.~Amati, M.~Ciafaloni and G.~Veneziano,
{\it Planckian scattering beyond the semiclassical approximation,}
  Phys.\ Lett.\  {\bf B289} (1992) 87.
  %%CITATION = PHLTA,B289,87;%%

  \bibitem{Peter} P. D. D'Eath and P. N. Payne, "Gravitational radiation in high speed black hole
collisions. 1. Perturbation treatment of the axisymmetric speed of light collision,"
Phys. Rev. D46 (1992) 658–674.\\
 P. D. D'Eath and P. N. Payne, "Gravitational radiation in high speed black hole
collisions. 2. Reduction to two independent variables and calculation of the second
order news function," Phys. Rev. D46 (1992) 675–693.\\
 P. D. D'Eath and P. N. Payne, "Gravitational radiation in high speed black hole
collisions. 3. Results and conclusions," Phys. Rev. D46 (1992) 694–701.

\bibitem{AVV}
I.Ya. Aref'eva, K.Viswanathan and I. Volovich, {\it Planckian-energy scattering,
  colliding plane gravitational waves and black hole creation},
  \NP {\bf B452} (1995) 346
 [Erratum-ibid.\  B {\bf 462}, 613 (1996)]
  hep-th/9412157
  %%CITATION = NUPHA,B452,346;%%
\\
   I.~Ya.~Arefeva, K.~S.~Viswanathan and I.~V.~Volovich,
  {\it On black hole creation in Planckian energy scattering},
  Int.\ J.\ Mod.\ Phys.\  D {\bf 5}, 707 (1996), hep-th/9512170
  %%CITATION = IMPAE,D5,707;%%




\bibitem{TeVgravity} N. Arkani-Hamed, S. Dimopoulos and G.R. Dvali, The Hierarchy Problem and New Dimensions
at a Millimeter, Phys.Lett. B429 (1998) 263, hep-ph/9803315;\\
I. Antoniadis, N. Arkani-Hamed, S. Dimopoulos and G.R. Dvali, New Dimensions at a Millimeter
to a Fermi and Superstrings at a TeV, Phys.Lett. B436 (1998) 257, hep-ph/9804398.

\bibitem{TeVgravity-HEC} G.F. Giudice, R. Rattazzi and J.D. Wells, {\it Quantum gravity and
extra dimensions at high-energy colliders}, Nucl. Phys. B 544 (1999)
3, hep-ph/9811291;
G.F. Giuduce, R. Rattazzi and J.D. Wells, Transplanckian Collisions at the LHC and Beyond,
Nucl.Phys. B630 (2002) 293, hep-ph/0112161.
\bibitem{BF}
T. Banks and W. Fischler, A Model for High Energy Scattering in Quantum Gravity,
hep-th/9906038
\bibitem{IA}
I.Ya. Aref'eva, High-energy scattering in the brane world and black hole production,
Part.Nucl.31 (2000) 169, hep-th/9910269
%%CITATION = PANUA,31,169;%%



\bibitem{DL} S. Dimopoulos and G. Landsberg, {\it Black Holes at the LHC},
Phys.Rev.Lett. 87 (2001) 161602,
hep-ph/0106295.
\bibitem{GT} S.B. Giddings and S.
Thomas, {\it High Energy Colliders as Black Hole Factories: The End of Short Distance
Physics}, Phys.Rev. D65 (2002) 056010,
 hep-ph/0106219
\bibitem{EG02} D.M. Eardley and S. B. Giddings, Classical Black Hole Production in High-Energy Collisions,
Phys.Rev. D66 (2002) 044011, gr-qc/0201034;
%%CITATION = PHRVA,D66,044011;%%
\bibitem{BH-list1} A. Ringwald, H. Tu, Collider versus cosmic ray sensitivity to black hole production,
Phys.Lett.B525 (2002) 135, hep-ph/0111042;\\
E.J. Ahn, M. Cavaglia and A.V. Olinto, Brane Factories, Phys.Lett. B551 (2003) 1,
hep-th/0201042;\\
S. N. Solodukhin, Classical and quantum cross-section for black hole production in particle
collisions, Phys.Lett. B533 (2002) 153; hep-ph/0201248.
\\
E.~Kohlprath and G.~Veneziano,
{\it Black holes from high-energy beam-beam collisions,}
  J. High Energy Phys {\bf 06} (2002) 057
{\tt  [arXiv:gr-qc/0203093].}
  %%CITATION = JHEPA,0206,057;%%
\bibitem{YN03}
H.~Yoshino and Y.~Nambu,
 {\it Black hole formation in the grazing collision of high-energy
particles}, \PR {\bf D67} (2003) 024009,
gr-qc/0209003
%%CITATION = GR-QC 0209003;%% YN03,YR05, num
\bibitem{YR05} H. Yoshino and V. S. Rychkov, Improved analysis of black hole formation in high-energy
particle collisions, Phys. Rev. D 71 (2005) 104028 ; hep-th/0503171.\\
M. Cavaglia, Black Hole and Brane Production in TEV Gravity: a Review, Int.J.Mod.Phys.
A18 (2003) 1843, hep-ph/0210296;\\
P.Kanti, Black Holes In Theories With Large Extra Dimensions: a Review, Int.J.Mod.Phys.
A19 (2004) 4899, hep-ph/0402168;\\
S.~B. Giddings and V.~S. Rychkov, {\it Black holes from colliding wavepackets},
  {\em Phys. Rev.} {\bf D70} (2004) 104026,
   hep-th/0409131.
\\
V. Cardoso, E. Berti and M. Cavaglia, What we (don�t) know about black hole formation in
high-energy collisions, Class.Quant.Grav.22:L61-R84,2005, hep-ph/0505125.\\
G.L. Landsberg, Black Holes at Future Colliders and Beyond, J.Phys.G32 (2006) R337,
hep-ph/0607297;\\
D.M. Gingrich, Black Hole Cross Section at the LHC, Int. J. Mod. Phys. A 21 (2006) 6653,
hep-ph/0609055;\\
H. Stoecker, Mini black holes in the first year of the LHC: Discovery through di-jet suppression,
multiple mono-jet emission and ionizing tracks in ALICE, J.Phys.G32 (2006) S429;\\
B. Koch, M. Bleicher, H. Stoecker, Black Holes at LHC?, J.Phys.G34 (2007) S535,
hep-ph/0702187;\\
N. Kaloper, J. Terning, How Black Holes Form in High Energy Collisions, arXiv:0705.0408.\\
M. Cavaglia et all., Signatures of black holes at the LHC, arXiv:0707.0317\\
P. Mende and L. Randall, {\it Black Holes and
Quantum Gravity at the LHC},
 arXiv:0708.3017\\
 S. B. Giddings, {\it High-energy black hole production}, arXiv:0709.1107.
 \bibitem{num}

  U.~Sperhake, V.~Cardoso, F.~Pretorius, E.~Berti and J.~A.~Gonzalez,
  %``The high-energy collision of two black holes,''
  Phys.\ Rev.\ Lett.\  {\bf 101}, 161101 (2008)
  [arXiv:0806.1738 [gr-qc]].
  %%CITATION = PRLTA,101,161101;%%

 M.~Shibata, H.~Okawa and T.~Yamamoto,
  %``High-velocity collision of two black holes,''
  Phys.\ Rev.\  D {\bf 78}, 101501 (2008)
  [arXiv:0810.4735 [gr-qc]].
  %%CITATION = PHRVA,D78,101501;%%


 U.~Sperhake, V.~Cardoso, F.~Pretorius, E.~Berti, T.~Hinderer and N.~Yunes,
  %``Cross section, final spin and zoom-whirl behavior in high-energy black hole
  %collisions,''
  Phys.\ Rev.\ Lett.\  {\bf 103}, 131102 (2009)
  [arXiv:0907.1252 [gr-qc]].
  %%CITATION = PRLTA,103,131102;%%


  M.~W.~Choptuik and F.~Pretorius,
  %``Ultra Relativistic Particle Collisions,''
  arXiv:0908.1780 [gr-qc].
  %%CITATION = ARXIV:0908.1780;%%








\bibitem{TM}
I.~Ya.~Aref'eva and I.~V.~Volovich,
  {\it Time Machine at the LHC},
  Int.\ J.\ Geom.\ Meth.\ Mod.\ Phys.\  {\bf 05}, 641 (2008),
  arXiv:0710.2696
  %%CITATION = 00436,05,641;%%

\bibitem{MMT}
  A.~Mironov, A.~Morozov and T.~N.~Tomaras,
  {\it If LHC is a Mini-Time-Machines Factory, Can We Notice?},
arXiv:0710.3395
  %%CITATION = 00271,4,381;%%
  \bibitem{NS}
  P.~Nicolini and E.~Spallucci,
  {\it Noncommutative geometry inspired wormholes and dirty black holes},
  arXiv:0902.4654.
  %%CITATION = ARXIV:0902.4654;%%

   \bibitem{INov}   I.~D.~Novikov, N.~S.~Kardashev and A.~A.~Shatskiy,
  ``The multicomponent Universe and the astrophysics of wormholes,''
  Phys.\ Usp.\  {\bf 50}, 965 (2007)
  [Usp.\ Fiz.\ Nauk {\bf 177}, 1017 (2007)].
  %%CITATION = UFNAA,177,1017;%%
\bibitem{NovNov} E.I. Novikova, I. D. Novikov, {\it Homogeneous singularities inside collapsing wormholes},
arXiv:0907.1936
  %%CITATION = ARXIV:0907.1936;%%

  %%%%%%%%%%%%%%%%%%%%%%%%%%%%%%
  \bibitem{adscft}  J.~M.~Maldacena,
  %``The large N limit of superconformal field theories and supergravity,''
  Adv.\ Theor.\ Math.\ Phys.\  {\bf 2}, 231 (1998)
  [Int.\ J.\ Theor.\ Phys.\  {\bf 38}, 1113 (1999)]
  [arXiv:hep-th/9711200].
  %%CITATION = HEP-TH 9711200;%%
  %%%%%%%%%%%%%%%%%%%%%%%%%%%%%%%%%
  %%%%%%SHOCK-plasma%%%%%%%%%
  %%%%%%%%%%%%%%%%
  \bibitem{Gubser08}
S.~S.~Gubser, S.~S.~Pufu and A.~Yarom,
{\it Entropy production in collisions of gravitational shock waves and of heavy
ions,}
  Phys.\ Rev.\  {\bf D78} (2008) 066014,
arXiv:0805.1551
  %%CITATION = PHRVA,D78,066014;%%
\bibitem{Alvarez08}
L.~Alvarez-Gaume, C.~Gomez, A.~S. Vera, A.~Tavanfar, and M.~A. Vazquez-Mozo,
  {\it Critical formation of trapped surfaces in the collision of gravitational
  shock waves}, arXiv:0811.3969

 \bibitem{Gubser09}
S.~S.~Gubser, S.~S.~Pufu and A.~Yarom,
 {\it Off-center collisions in AdS with applications to multiplicity
 estimates in heavy-ion collisions},
 arXiv:0902.4062 [hep-th]

\bibitem{0803.1467} D. ~M. ~Hofman, J.~M. ~Maldacena, ``Conformal collider physics: Energy and charge correlations,''
 JHEP \textbf{0805} (2008) 012, arXiv:0803.1467 [hep-th]
\bibitem{Shuryak08}
  E.~Shuryak,
  ``Physics of Strongly coupled Quark-Gluon Plasma,''
  Prog.\ Part.\ Nucl.\ Phys.\  {\bf 62}, 48 (2009)
  [arXiv:0807.3033 [hep-ph]].
  %%CITATION = PPNPD,62,48;%%
\bibitem{0803.3226} D.~Grumiller, P.~Romatschke, ``On the collision of two shock waves in AdS5,''
JHEP \textbf{0808} (2008) 027, arXiv:0803.3226 [hep-th]


\bibitem{Shuryak09}
  S.~Lin and E.~Shuryak,
  ``Grazing Collisions of Gravitational Shock Waves and Entropy Production in Heavy Ion Collision,''
  arXiv:0902.1508 [hep-th].
  %%CITATION = ARXIV:0902.1508;%%

\bibitem{0902.3046} J.~L. ~Albacete, Y.~V. ~Kovchegov, A. ~Taliotis,
``Asymmetric Collision of Two Shock Waves in $AdS_5$,'' JHEP \textbf{0905} (2009) 060, arXiv:0902.3046 [hep-th]
\bibitem{0904.2536} W.~A. ~Horowitz, Y.~V. ~Kovchegov, ``Shock Treatment: Heavy Quark Drag in a Novel AdS Geometry,'' arXiv:0904.2536 [hep-th]
\bibitem{0906.4426} P.~M. ~Chesler, L.~G. ~Yaffe, ``Boost invariant flow, black hole formation, and far-from-equilibrium dynamics in N = 4 supersymmetric Yang-Mills theory,'' arXiv:0906.4426 [hep-th]
\bibitem{0907.4604} E. ~Avsar, E. ~Iancu, L. ~McLerran, D.~N. ~Triantafyllopoulos, ``Shockwaves and deep inelastic scattering within the gauge/gravity duality,'' arXiv:0907.4604 [hep-th]
\bibitem{0908.3677} G.~T. ~Horowitz, M.~M. ~Roberts, ``Zero Temperature Limit of Holographic Superconductors,''
arXiv:0908.3677 [hep-th]
  %%%%%%%%%%%%%%%%%%%%
  %%%%%%plasma-ads/cft-equilibrium
    \bibitem{Tseytlin} S.~S.~Gubser, I.~R.~Klebanov and A.~A. ~Tseytlin, {Nucl.\ Phys.\ } {\bf B534} (1998) 202

\bibitem{Starinets}G.~Policastro, D.~T.~Son and A.~O.~Starinets,
``The shear viscosity of strongly coupled N = 4 supersymmetric Yang-Mills  plasma,``
Phys.\ Rev.\ Lett.\  {\bf 87} (2001) 081601.
%CITATION = HEP-TH 0104066;%%
\bibitem{Solana} J.~Casalderrey-Solana and D.~Teaney,
``Heavy quark diffusion in strongly
  coupled N = 4 Yang Mills,''
 {{\tt hep-ph/0605199}}.
\bibitem{Zahed} S.-J. Sin and I.~Zahed,
 ``Holography of radiation and jet quenching,''
 {\em
  Phys. Lett.} {\bf B608} (2005) 265--273,
  {{\tt hep-th/0407215}}.\\
H.~Liu, K.~Rajagopal, and U.~A. Wiedemann,
 ``Calculating the jet quenching
 parameter from AdS/CFT,''
  hep-ph/0605178.\\
C.~P. Herzog, A.~Karch, P.~Kovtun, C.~Kozcaz, and L.~G. Yaffe,
``Energy loss of
  a heavy quark moving through N = 4 supersymmetric Yang-Mills
plasma,''
{{\tt hep-th/0605158}}.\\
S.~S. Gubser, ``Drag force in AdS/CFT,''Phys.Rev.D74:126005,2006,  {{\tt hep-th/0605182}} \\
A.~Buchel,
`On jet quenching parameters in strongly coupled non- conformal
  gauge theories,''
 {{\tt
  hep-th/0605178}}.
 {{\tt hep-th/0605182}}.\\
S.-J. Sin and I.~Zahed,
``Ampere's Law and Energy Loss in AdS/CFT Duality,''
  {{\tt hep-ph/0606049}}.
  S. Bhattacharyya, V. E. Hubeny, S. Minwalla, and M. Rangamani, "Nonlinear Fluid
Dynamics from Gravity," JHEP 02 (2008) 045, 0712.2456.

  %%%%%%%%%%%%%%
  \bibitem{Brill:1993tw} D.~R.~Brill and S.~A.~Hayward, ``Global structure of
a black hole cosmos and its extremes,'' Class.\ Quant.\ Grav.\ \textbf{11}
(1994) 359 [arXiv: gr-qc/9304007].
%%CITATION = GR-QC 9304007;%%

\bibitem{Romans:1991nq} L.~J.~Romans, ``Supersymmetric, cold and lukewarm
black holes in cosmological Einstein-Maxwell theory,'' Nucl.\ Phys.\ B
\textbf{383} (1992) 395 [arXiv: hep-th/9203018].
%%CITATION = HEP-TH 9203018;%%

\bibitem{ross} S.~F.~Ross and R.~B.~Mann, ``Cosmological production of
charged black hole pairs,`` Phys.\ Rev.\ D \textbf{52} (1995) 2254 [arXiv:
gr-qc/9504015].
%%%%%%%%%%%%%%%%%%%%%%%%%%%%%%%%%%%%%%%%%%%%%%%%%%%%%%%%%%%%%%%%%%%%%%%%%%%%%%

\bibitem{bousso} R.~Bousso, ``Quantum global structure of de Sitter space''
Phys.\ Rev.\ D \textbf{60} (1999) 063530 [arXiv: hep-th/9902183].
%%CITATION = 9902183;%%
\bibitem{Jolien} J. D. E. Creighton and R. B. Mann, ``Quasilocal
thermodynamics of dilaton gravity coupled to gauge fields,'' Phys.\ Rev. \ D
\textbf{52} (1995) 4569 [arXiv: gr-qc/9505007].
%%CITATION = GR-QC 9505007;%%
\bibitem{Kastor} D. Kastor, J. Traschen, {\it Cosmological multi-black-hole solutions}, Phys. Rev. D, vol. 47, ¹ 12 (1993).



  %%%%%%%%%%%%%%%%%%%%

\bibitem{Narayan:2005ie}
  R.~Narayan,
  {\it Black Holes in Astrophysics},
  New J.\ Phys.\  {\bf 7}, 199 (2005),
  gr-qc/0506078
  %%CITATION = NJOPF,7,199;%%

  %%%%%%%%%%%%%%%% DS/CFT
  \bibitem{Hull98} C.~M.~Hull, ``Timelike T-duality, de Sitter space,
large N gauge theories and topological field theory,'' JHEP \textbf{9807}
(1998) 021 [arXiv: hep-th/9806146]. %%CITATION = HEP-TH 9806146;%%

\bibitem{Hull98signature} C.~M.~Hull, ``Duality and the signature of
space-time,'' JHEP \textbf{9811} (1998) 017 [arXiv: hep-th/9807127].
%%CITATION = HEP-TH 9807127;%%

\bibitem{Minic01} V.~Balasubramanian, P.~Horava and D.~Minic,
``Deconstructing de Sitter,'' JHEP \textbf{0105} (2001) 043 [arXiv:
hep-th/0103171]. %%CITATION = HEP-TH 0103171;%%

\bibitem{Witten01ds-dul} E.~Witten, ``Quantum gravity in de Sitter space,''
arXiv:hep-th/0106109. %%CITATION = HEP-TH 0106109;%%

\bibitem{Strominger01ds-dul} A.~Strominger, ``The dS/CFT correspondence,''
JHEP \textbf{0110} (2001) 034 [arXiv: hep-th/0106113].
%%CITATION = HEP-TH 0106113;%%
%%%%%%%%%%%%%%%%%%%%%%%%%%%%

\bibitem{YosMann} H.Yoshino and R. B. Mann, {\it Black hole formation in the head-on collision
 of ultrarelativistic charges}, 0605131



\bibitem{Hatta-charge} Y. Hatta, {\it  $e^+e^-$ annihilation to high energy scattering at weak and
strong coupling},
JHEP0811:057,2008, arXiv:0810.0889

\bibitem{giddings}
  D.~M.~Eardley and S.~B.~Giddings,
  %``Classical black hole production in high-energy collisions,''
  Phys.\ Rev.\  D {\bf 66}, 044011 (2002)
  [arXiv:gr-qc/0201034].
  %%CITATION = PHRVA,D66,044011;%%

  \bibitem{nambu}
  H.~Yoshino and Y.~Nambu,
  %``Black hole formation in the grazing collision of high-energy particles,''
  Phys.\ Rev.\  D {\bf 67}, 024009 (2003)
  [arXiv:gr-qc/0209003].
  %%CITATION = PHRVA,D67,024009;%%


\bibitem{veneziano}
  E.~Kohlprath and G.~Veneziano,
  %``Black holes from high-energy beam-beam collisions,''
  JHEP {\bf 0206}, 057 (2002)
  [arXiv:gr-qc/0203093].
  %%CITATION = JHEPA,0206,057;%%

\bibitem{ABG} I.Ya. Aref'eva, A.A. Bagrov, E.A. Guseva, {\it Critical Formation of Trapped Surfaces
in the Collision of Non-expanding Gravitational Shock Waves in de Sitter Space-Time}, arXiv:0905.1087.\\
I.Ya. Aref'eva, A.A. Bagrov, {\it Trapped sufaces formation in collisions of non-expanding gravitational
shock waves in $AdS_4$ spacetime}, Theor. Math. Phys., vol. 161, ¹3, 385-403 (2009)



%%%%%%%%%%%%%%%%%%%%%OLD papers

\bibitem{LS90} C.~O.~Lousto and N.~Sanchez,
{\it The Curved Shock Wave Space-Time Of Ultrarelativistic Charged Particles And
Their Scattering},
Int.\ J.\ Mod.\ Phys.\ A \textbf{5}, 915 (1990),
{\it Scattering Processes at the Planck Scale Authors}, gr-qc/9410041
%%CITATION = IMPAE,A5,915;%%
\bibitem{Ortaggio06}
M.~Ortaggio,
``Ultrarelativistic boost of spinning and charged black rings,''
arXiv:gr-qc/0601093.
\bibitem{AS71}
P.~C.~Aichelburg and R.~U.~Sexl,
``On the gravitational field of a massless particle,''
Gen.\ Rel.\ Grav.\  {\bf 2}, 303 (1971).
%%CITATION = GRGVA,2,303;%%



\bibitem{MP86}
R.~C.~Myers and M.~J.~Perry,
``Black Holes In Higher Dimensional Space-Times,''
Annals Phys.\  {\bf 172}, 304 (1986).
%%CITATION = APNYA,172,304;%%



\bibitem{Gibbons-Kerr} G. W. Gibbons, H. Lu, D. N. Page, and C. N. Pope,
{\it The general Kerr-de Sitter metrics
in all
dimensions}, J. Geom. Phys. 53 (2005) 49–73, hep-th/0404008.
%%%%%%%%%%%%%%% SW-curve  HT,LS90,Sfetsos,GrifPod,GrifPod,Emparan,Ortaggio06,Eso
\bibitem{HT}
Hotta~M., Tanaka~M. {\it Shock wave geometry with non-vanishing
  cosmological constant}, \CQG {\bf 10} (1993) 307.




\bibitem{Sfetsos}  K. Sfetsos, "On gravitational shock waves in curved space-times," Nucl. Phys. B436
(1995) 721–746, hep-th/9408169.
hep-th/9408169
\bibitem{GrifPod} J. Podolsky and J. B. Griffiths, "Impulsive waves in de Sitter and anti-de Sitter
space-times generated by null particles with an arbitrary multipole structure," Class.
Quant. Grav. 15 (1998) 453–463, gr-qc/9710049.

\bibitem{Emparan} R. Emparan, "Exact gravitational shockwaves and Planckian scattering on branes,"
Phys. Rev. D64 (2001) 024025, hep-th/0104009.

%%CITATION = GR-QC 0601093;%%
\bibitem{Eso} Esposito G., Pettorino R., Scudellaro P., On boosted space-times with cosmological
constantand their ultrarelativistic limit, arXiv:gr-qc/0606126v3.


%%%%%%%%%%%%%%%%%%%%%%
\bibitem{Nastase1} K. Kang and H. Nastase, "Planckian scattering effects and black hole production in
low M(Pl) scenarios," Phys. Rev. D71 (2005) 124035, hep-th/0409099.
\bibitem{Nastase2} K. Kang and H. Nastase, "High energy QCD from Planckian scattering in AdS and
the Froissart bound," Phys. Rev. D72 (2005) 106003, hep-th/0410173.
\bibitem{Nastase3}H. Nastase, "On high energy scattering inside gravitational backgrounds,"
hep-th/0410124.

\bibitem{Radu} D. Astefanesei, R. B. Mann and E. Radu,
{\it  Reissner-Nordstrom-de Sitter black hole, planar coordinates and dS / CFT},
JHEP 0401:029,2004, hep-th/0310273





Y. Brihaye, T. Delsate, {\it  Charged-Rotating Black Holes in Higher-dimensional
(A)DS-Gravity},
arXiv:0806.1583





%%%%%%%%%%%%%%about general theorem conserning BH production
\bibitem{HawkingPenrose}S. W. Hawking and R. Penrose, "The Singularities of gravitational collapse and
cosmology," Proc. Roy. Soc. Lond. A314 (1970) 529–548.
\bibitem{Penrose} R. Penrose, "The question of cosmic censorship," J. Astrophys. Astron. 20 (1999)
233–248.
 %%%%%%%%%%%%%%%%%%%%%%%
%%%%%%%%%%%%%%%%%%about tS->bh in AdS Gibbons,0001003,0004032,0803.2526
\bibitem{Gibbons}G. W. Gibbons, "Some comments on gravitational entropy and the inverse mean
curvature flow," Class. Quant. Grav. 16 (1999) 1677–1687, hep-th/9809167.
\bibitem{0001003} P. T. Chrusciel, E. Delay, G. J. Galloway, and R. Howard, "The Area Theorem,"
Annales Henri Poincare 2 (2001) 109–178, gr-qc/0001003.
\bibitem{0004032} P. T. Chrusciel and W. Simon, "Towards the classification of static vacuum spacetimes
with negative cosmological constant," J. Math. Phys. 42 (2001) 1779–1817,
gr-qc/0004032.
\bibitem{0803.2526}S. Bhattacharyya et. al., "Local Fluid Dynamical Entropy from Gravity," 0803.2526.


%%%%%%%%%%%%%%%%%%%%%%%%%%%%%%%%%%%%%%%%%%%%%%%%%%%%%%%%%%%%%%%%%%%%%%%%%%%%%%
\bibitem{Gel} I.M. Gelfand, G.E. Shilov, {\it Generalized functions}, vol. 1, 2,
 New York and London, Academic Press Inc, 1964
\bibitem{VSV} V.S.Vladimirov, {\it Generalized functions in mathematical physics}, Moscow, 1976, (In Russian);
English transl.("Mir", Moscow), 1979

\bibitem{hoop} T. Chiba, K. Maeda, {\it Cosmic hoop conjecture?}, Phys. Rev. D, vol. 50, ¹ 8 (1994).\\
D. Ida, K. Nakao, M. Siino, S. Hatward, {\it Hoop conjecture for colliding black holes}, Phys. Rev. D {\bf 58} 121501
(1998)

\end{thebibliography}
  \end{document}